\documentclass[fleqn,10pt,twocolumn]{wlscirep}
\usepackage[utf8]{inputenc}
\usepackage[T1]{fontenc}

\newcommand{\beq}{\begin{eqnarray}}
\newcommand{\eeq}{\end{eqnarray}}
\usepackage{amsmath}
\usepackage{tikz}
\usepackage{balance}
\usepackage{soul}
\usepackage{color}
\usetikzlibrary{decorations.pathmorphing}
\usetikzlibrary{shapes.misc}
\tikzset{cross/.style={cross out, draw=black, minimum size=8*(#1-\pgflinewidth), inner sep=0pt, outer sep=0pt},
cross/.default={1pt}}
\usetikzlibrary{patterns,math}
\usepackage[normalem]{ulem}
\usepackage[columnwise]{lineno}
\usepackage{verbatim}
\usepackage{color,ulem}
\usepackage[english]{babel}
\usepackage{caption}

\definecolor{purple}{rgb}{0.41, 0.16, 0.38}

\title{Non-linear elasticity, yielding and entropy in amorphous solids}

\makeatletter
\def\thanks#1{\protected@xdef\@thanks{\@thanks\protect\footnotetext{#1}}}
\makeatother

\author[1,$\mathsection$]{Deng Pan}
\author[1,2,$\mathsection$]{Teng Ji}
\author[3,$^\ast$\thanks{$^\ast$e-mail: b.matteo@sjtu.edu.cn}]{Matteo Baggioli}
\author[1,2,4,$^\ast$\thanks{$^\ast$e-mail: liliphy@itp.ac.cn}]{Li Li}
\author[1,2,5$^\ast$\thanks{$^\ast$e-mail: yuliangjin@mail.itp.ac.cn}]{Yuliang Jin}
\affil[1]{CAS Key Laboratory of Theoretical Physics, Institute of Theoretical Physics,
Chinese Academy of Sciences, Beijing 100190, China}
\affil[2]{School of Physical Sciences, University of Chinese Academy of Sciences, Beijing 100049, China}
\affil[3]{Wilczek Quantum Center, School of Physics and Astronomy, Shanghai Jiao Tong University, Shanghai 200240, China \& Shanghai Research Center for Quantum Sciences, Shanghai 201315, China}
\affil[4]{School of Fundamental Physics and Mathematical Sciences, Hangzhou Institute for Advanced Study, University of Chinese Academy of Sciences, Hangzhou 310024, China}
\affil[5]{\color{black} Wenzhou Institute, University of Chinese Academy of Sciences, Wenzhou, Zhejiang 325000, China \color{black}}

\affil[$\mathsection$]{These authors contributed equally}

\begin{abstract}
The holographic duality has proven successful in linking seemingly unrelated problems in physics.
Recently, intriguing correspondences between the physics of soft matter and gravity are emerging,
including strong
similarities between the rheology of amorphous solids, effective field theories for elasticity and the physics of black holes.
However, direct comparisons between theoretical predictions and experimental/simulation  observations remain limited.
Here, we study the effects of non-linear elasticity on the mechanical and thermodynamic properties of amorphous materials responding to shear, using effective field and gravitational theories. The predicted correlations among the non-linear elastic exponent, the yielding strain/stress and the entropy change due to shear are supported qualitatively  by simulations of granular matter models. Our approach  opens a path towards understanding complex mechanical responses of amorphous solids, such as mixed effects of shear softening and shear hardening, and offers the possibility to study the rheology of solid states and black holes in a unified framework.
\end{abstract}
\begin{document}

\flushbottom
\maketitle

\thispagestyle{empty}

\section*{Introduction}

Amorphous solids, including but not limited to, glasses, granular matter, colloidal suspensions, foams and polymers,  yield to external shear~\cite{bonn2017yield, nicolas2018deformation}. The yielding point, typically characterized by a maximum on the stress-strain curve, is essentially the end-point of the solid regime, as the material transitions to  plastic flow thereafter.
Understanding the nature of yielding in amorphous solids, in the context of statistical mechanics, has been becoming an active research area. Recent theories attempt to explain yielding as a depinning transition~\cite{nicolas2018deformation}, a first-order non-equilibrium phase transition~\cite{jaiswal2016mechanical, kawasaki2016macroscopic}, a spinodal point of the glass state of equation~\cite{rainone2015following, parisi2017shear}, or phase transitions in the universality class of a random field Ising model~\cite{ozawa2018random}. However, treating non-linear effects is intellectually challenging~\cite{hentschel2011athermal, biroli2016breakdown}, and consequently their impact on yielding remains unclear to a large extent.

Non-linear elastic responses, in particular {\it shear hardening}, as evidenced by a rapid increase of the shear modulus before yielding, are not uncommon in amorphous solids. Shear hardening has been widely observed in polymers~\cite{treloar1975physics}, and more recently  in dense hard sphere (HS) colloidal glasses close to jamming~\cite{jin2018stability, urbani2017shear}. In both cases, hardening is accompanied by entropy vanishing caused by structural constraints:
the number of allowed configurations tends to zero approaching the maximum stretch limit in polymer chains, and the jamming limit in HSs. In this study, we discover a new type of shear hardening in amorphous solids, where the entropy increases with strain, caused by shear-induced rejuvenation. This non-linearity also leads to  a negative correlation between the yielding strain and the degree of annealing, in sharp contradiction to previous results~\cite{rainone2015following, jin2017exploring, ozawa2018random}.

\begin{figure}[ht]
    \centering
    \includegraphics[width=0.9\linewidth]{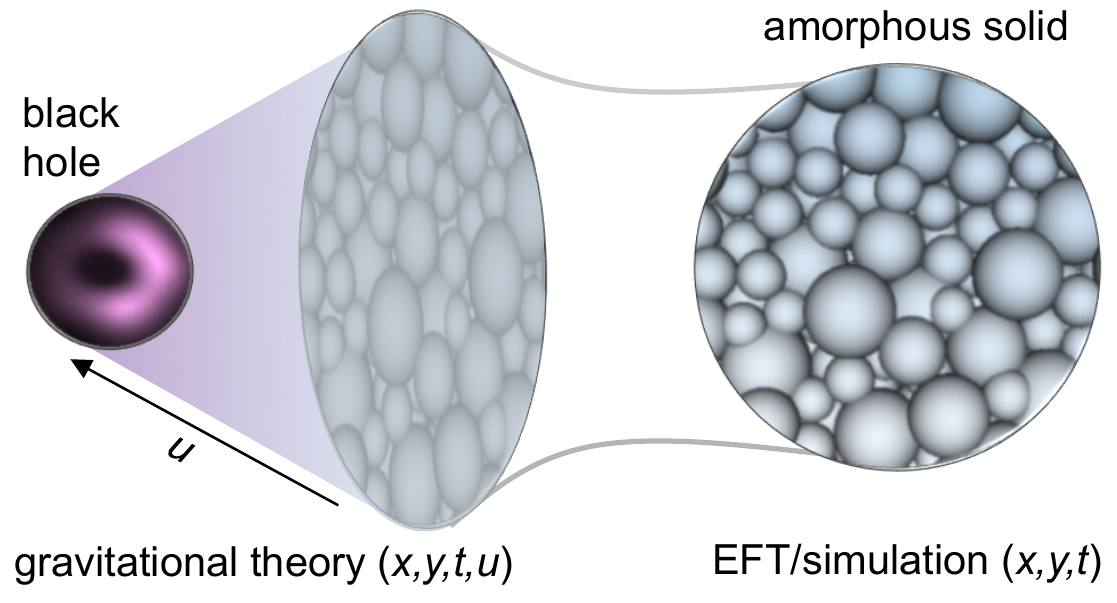}
   \caption{\textbf{A schematic representation of the holographic duality.} The gravitational models live in $(3+1)$ dimensions while effective field theories/amorphous solid simulations are in $(2+1)$ dimensions.}
    \label{cart}
\end{figure}

Let us sketch out our strategy. (i)
First, a zero-temperature effective field theory (EFT) for non-linear elasticity is constructed, based on shift-symmetric Goldstone fields, the phonons, as fundamental building blocks~\cite{PhysRevD.102.069901}. The theory also takes advantage of the fact that amorphous solids are typically rotationally invariant (isotropic), and assumes that the solid is homogeneous at low energy, which is true for systems at large spatial scales and/or slow dynamics, such as granular matter under quasi-static shear. The non-linear elasticity is implemented in the form of an effective potential (cf. strain energy function), from which the stress-strain curve, as well as the onset of instability (yielding), can be calculated.

(ii) To  overcome the difficulty in treating dissipative (finite temperature) effects in the EFT framework, we map the EFT, defined in a $(2+1)-$dimensional flat spacetime with coordinates $(x, y, t)$, onto a gravitational dual model~\cite{Baggioli:2020qdg} in asymptotically anti-de Sitter spacetime $(x, y, t, u)$, according to the  holographic duality (or gauge-gravity duality) ~\cite{Ammon:2015wua, Baggioli:2019rrs, Roy:2020vku,PhysRevLett.120.171602, PhysRevLett.120.195301}, where the extra dimension $u$ represents the energy scale (see Fig.~\ref{cart}). The computation of the entropy becomes particularly simple now, since it boils down to the estimation of the black hole entropy given by the famous Bekenstein - Hawking Area law~\cite{hawking1971gravitational}. Such a holographic approach has been already successfully applied to study the dynamics of strongly coupled fluids (e.g., Quark-Gluon plasma) \cite{CasalderreySolana:2011us} and novel strongly correlated materials (e.g., Cuprates) \cite{Hartnoll:2016apf}.

(iii) To test the predictions from the EFT and the gravitational theory, we perform numerical simulations on soft sphere (SS) models of granular matter. Consistent with the theoretical setup, shear is applied quasi-statically, thermal motions are neglected (for macroscopic grains, the thermal energy is much smaller than  the inter-particle contact energy), and plastic effects are irrelevant. We consider a common type of shear, planar shear (such as simple or pure shear), where the material is unaffected along the dimensions perpendicular to the plane and the effective rheology is two dimensional (see Supplementary Materials (SM) Sec.~S6). The spatial dimensionality is thus reduced to $d=2$ in the theories. The simulations are performed in both 3D (main text) and 2D (SM { Secs.~S7 and S8}). Below, we discuss our results in detail.

\begin{figure*}[ht!]
    \centering
      \includegraphics[width=\linewidth]{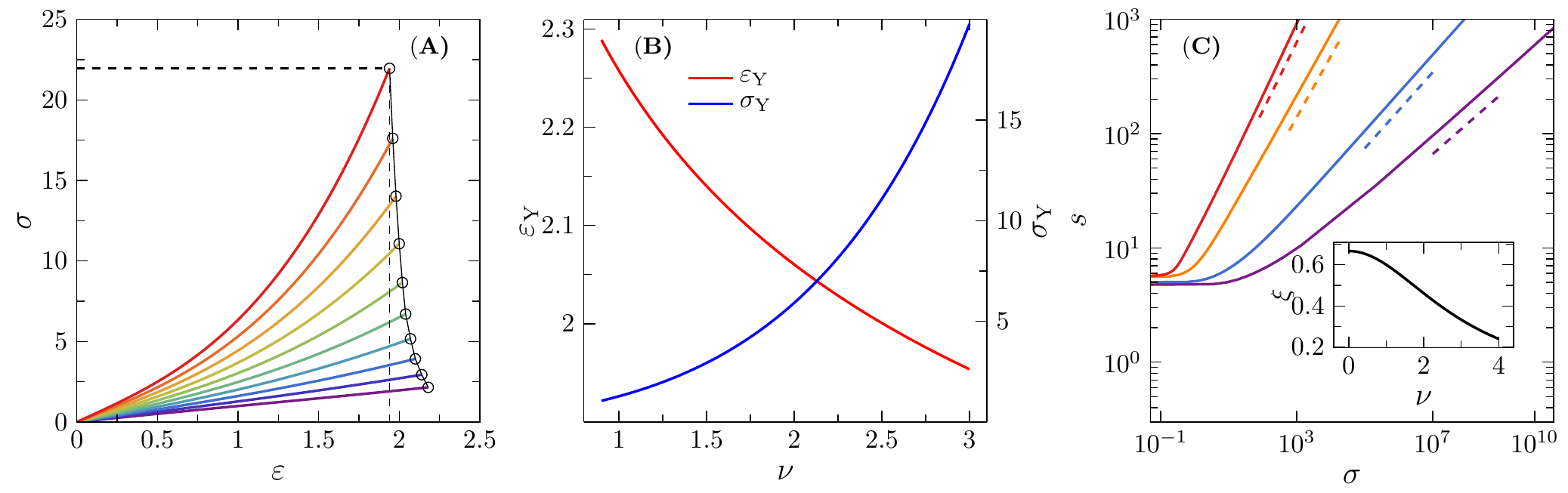}
    \caption{\textbf{Predictions from the EFT and the gravity theory}.
    \textbf{(A)} Stress-strain curves extracted from the EFT. The Poisson's ratio is fixed to $\mathfrak{r} = 97\%$, nearly the limit of incompressibility. The non-linear shear exponent $\nu$ is varied from $1.3$ to $3.1$ (from \textbf{purple} to \textbf{red}) in steps of $0.2$. The empty circles locate the breaking points. \textbf{(B)} The relation between the non-linear exponent $\nu$ and the  { breaking point} strain $\epsilon_{\rm Y}$ (\textbf{red}) or the {breaking point} stress $\sigma_{\rm Y}$ (\textbf{blue}) for $\mathfrak{r}=97\%$. \textbf{(C)} The entropy-stress curves for $\nu = 1.00, 1.50, 3.00, 3.75$ (from \textbf{purple} to \textbf{red}),
    from the gravity theory at $T=0.1$, with the large stress scaling Eq.~(\ref{eq:s0}) indicated (dashed lines). \textbf{(Inset)} The theoretical relation $\xi = 2/(3+\frac{1}{3}\nu^2)$.}
    \label{fig:max}
\end{figure*}

\section*{Results}

\subsection*{Correlation between yielding and non-linear elasticity predicted by an EFT}

The EFT builds on two scalar fields $   \phi^I(t,\bf{x})$, whose fluctuations $\delta \phi^I(t,\bf{x})$ represent the displacement fields in the solid ($I=1,2$). The stress-strain curves are obtained from a power-law scalar potential $\mathcal{V}$, which plays the same role of the non-linear strain-energy function in standard elasticity theories (see Materials and Methods and SM Secs.~S1 and S2 for details). The assumptions used in the theory include: (i) rotational invariance (amorphicity), (ii) homogeneity at large scales (larger than the granularity scale of the solid), and (iii) absence of plastic effects. The validity of these assumptions is tested by a direct comparison between theoretical and simulation results.

The only input to the theory is a power-law form of the stress-strain relation (see Fig.~\ref{fig:max}(A)),
\beq
\sigma(\epsilon)\,\sim \,\epsilon^{\nu},
\label{eq:nonlinear_scaling1}
\eeq
which corresponds to a concrete choice of the potential $\mathcal{V}$. Here, $\nu = 2A$ is an exponent characterizing the non-linear elasticity, which cannot be determined directly from the effective theory, but rather should be considered  as a phenomenological parameter. A complete microscopic model, able for example to describe the dependence on  initial conditions, would be needed in order to determine directly such an exponent from theory. To continue, an important quantity defined in the theory is the Poisson's ratio (the negative ratio of transverse to axial strains),
\beq
\mathfrak{r}=\frac{B(B-1)-A}{B(B-1)+A},
\label{poisso1}
\eeq
which is related to the unstrained bulk modulus $K_0$ and the shear modulus $G_0$ via $\mathfrak{r} = (K_0-G_0)/(K_0+G_0)$ in two dimensions (notations with subscript $_0$ are defined at zero strain). Because there are only two independent parameters $A$ and $B$ (see Materials and Methods), the model is completely fixed by a given combination of $\nu$ and $\mathfrak{r}$.

An interesting prediction is the dependence of {\it breaking point} $\{ \epsilon_{\rm Y}, \sigma_{\rm Y}\}$, defined by a global instability where the speed of sound vanishes, on the non-linear elasticity. Since the theory is constructed in the hydrodynamic (long-wavelength) limit, this instability is naturally associated with the yielding of the whole system, instead of local plastic rearrangements. Analytic formulas for $ \epsilon_{\rm Y}$ and $\sigma_{\rm Y}$ are derived and  presented in SM Sec.~S2. For a fixed, large $\mathfrak{r}$ (i.e., $K_0 \gg G_0$), corresponding to a nearly incompressible material, $\nu$  is negatively correlated  with $\epsilon_{\rm Y}$, and
positively correlated with $\sigma_{\rm Y}$ (see Fig.~\ref{fig:max}(B) and SM Sec.S2 for more details). Moreover, in the case of a power-law form of the stress-strain curve as in Eq.~\eqref{eq:nonlinear_scaling1}, the EFT predicts a power-law correlation between the
{breaking point} strain $\epsilon_{\rm Y}$ and the non-linear exponent $\nu$ of the type $\epsilon_{\rm Y}\sim \nu^{-\kappa}$, where $\kappa>0$ depends on the specific details of the EFT potential (see SM Sec. S2). Thus, non-linear elasticity increases the strength (maximum stress) of the material, while making it more brittle and having an earlier breaking point.

\subsection*{Scaling of entropy predicted by a gravitational theory}

In the holographic description~\cite{Baggioli:2020qdg}, gravity is coupled to a scalar potential $W$, which depends on two bulk fields $\varphi^I$ (see Materials and Methods and SM Sec.~S3 for details).
For a given shear strain $\epsilon$, the stress $\sigma$ is read off from the black hole geometry using the holographic dictionary, and a power-law scaling as in Eq.~\eqref{eq:nonlinear_scaling1} is recovered (see SM Sec.~S4).
The breaking point can also be obtained by looking at the gradient instabilities of the gravitational modes and, at least in the decoupling limit, its behaviour as a function of the non-linear exponent $\nu$ is identical to the EFT results reported above~\cite{Baggioli:2020qdg}. Let us emphasize that the power-law exponent $\kappa$ and even the power-law functional form are not expected to be universal but rather dependent  on the details of the concrete model. Nevertheless, in all cases investigated $\varepsilon_{\rm Y}$ (or $\sigma_{\rm Y}$) is a monotonically decreasing (or increasing) function of the non-linear exponent $\nu$.
Therefore, we conclude that the correlations between the breaking point and the non-linear exponent are robust at the qualitative level.

The zero-temperature entropy $s_{\text{ath}} = s(T \to 0)$, computed from the theory, remains non-zero\cite{PhysRevX.5.041025}, which surprisingly resembles a key feature of amorphous solids. It is well known that the total entropy of an amorphous solid can be decomposed into configurational and vibrational parts, $s = s_{\rm conf} + s_{\rm vib}$.
When $T \to 0$,
the vibrational entropy $s_{\rm vib} \to 0$, while the configurational entropy $s_{\rm conf}$ and the total entropy $s$ remain finite, in contrast to crystals where $s \to 0$.  The finite $s_{\text{ath}}$ indicates the possibility of  multiple meta-stable states, revealing the glassy nature of black hole systems \cite{DeGiuli:2020zew,PhysRevB.100.205108, PhysRevX.5.041025}. As an essential result from our theory, the zero-temperature entropy, $s_{\text{ath}} \sim s_{\rm conf}$, scales with the shear stress $\sigma$ as (for large $\sigma$),
\beq
s_\text{ath} \sim \sigma^{\xi}\,,
\label{eq:s0}
\eeq
where the exponent $\xi = 2/(3+\nu^2/3)$. The prefactor vanishes at $T=0$, but the scaling is robust at sufficiently low temperatures (see SM Sec.~S5). Since $\xi$ is positive, the entropy always increases under shear, implying a rejuvenation effect -- the system becomes  more entropic and less stable under shear. This effect is overlooked by state-following (SF) calculations in mean-field glass theories~\cite{rainone2015following}, where, by construction, the entropy is kept constant during shear. Importantly, the increase of entropy under shear deformation is in stark contrast with the behavior observed in ordered crystals: e.g. high-density face-centered cubic HS crystals jam, and therefore their (vibrational) entropy vanishes (the configurational entropy $s_{\rm conf} = 0$ in crystals), under shear~\cite{jin2021jamming}. It suggests that our  holographic models might share more commonalities with amorphous  rather than crystalline systems, in agreement with recent related considerations \cite{PhysRevB.100.205108}.

\begin{figure*}[ht!]
    \centerline{    \includegraphics[width=\linewidth]{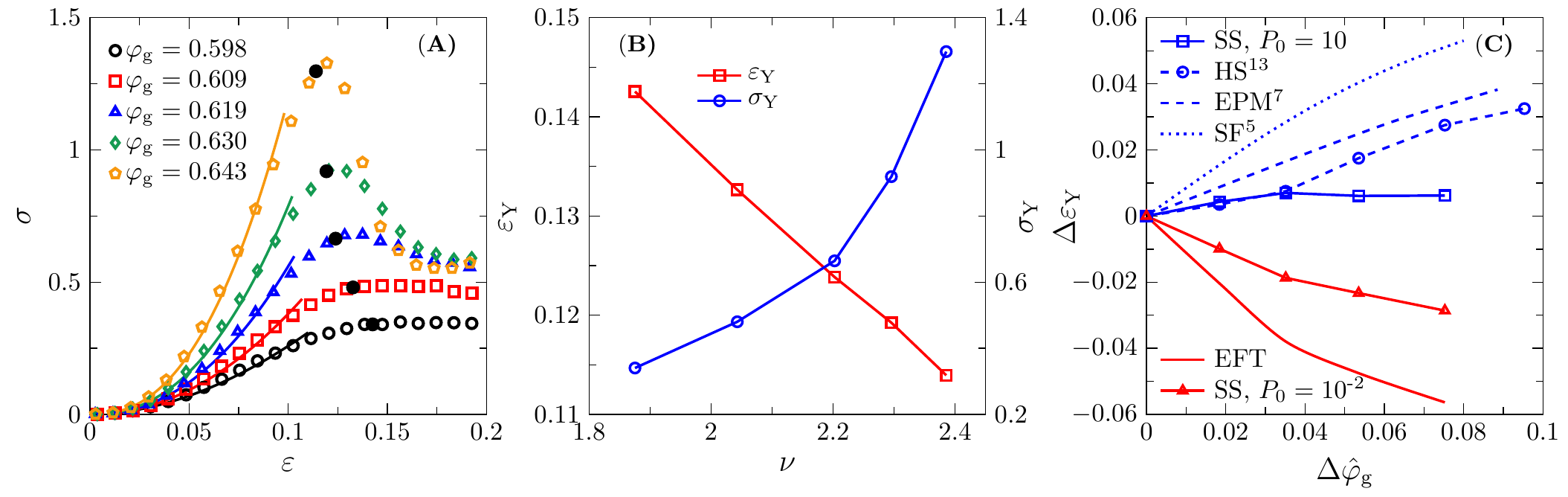}}
    \caption{\textbf{Correlation between non-linear elasticity and yielding obtained from granular simulations.}
    \textbf{(A)} Stress-strain curves for $P_{0}=10^{-2}$ and a few different $\varphi_{\rm g}$. The yielding point $\{ \epsilon_{\rm Y}, \sigma_{\rm Y}\}$ is estimated  at $\sigma_{\rm Y} = c \sigma_{\rm max}$, where $\sigma_{\rm max}$ is the maximum stress and $c = 0.98$ (see SM Sec.~S6 for other choices of $c$ and the discussion therein). Fitting the data (solid lines) according to Eq.~(\ref{eq:nonlinear_scaling1}) gives the exponent $\nu$ (see SM Sec.~S6 for a discussion on the fitting).
    \textbf{(B)} $\epsilon_{\rm Y}$ and $\sigma_{\rm Y}$ as functions of $\nu$.
    \textbf{(C)} Dependence of $\Delta \epsilon_{\rm Y} = \epsilon_{\rm Y}(\Delta \hat{\varphi}_{\rm g}) - \epsilon_{\rm Y}(0) $ on re-scaled degree of annealing $\Delta \hat{\varphi}_{\rm g} = (\varphi_{\rm g} - \varphi_{\rm MCT})/ \varphi_{\rm MCT}$, where  $\varphi_{\rm MCT}$ is the  mode-coupling theory (MCT) transition density.
    Besides EFT and simulation results  of athermal SSs ($P_{0}=10$ and $10^{-2}$) obtained in this study, we present as well simulation data of 3D thermal HSs~\cite{jin2017exploring} and  the SF theoretical result (divided by 5)~\cite{rainone2015following}. In addition, we plot the theoretical result from the elasto-plastic model (EPM), where $x$-axis represents $(A_{\rm c} - A)/A$ with $A$ being the degree of annealing and $A_{\rm c}$ a critical point~\cite{ozawa2018random}. Theories are indicated by lines and simulations by line-points. It is clear that  $\epsilon_{\rm Y}$ decreases with the degree of annealing due to shear-hardening non-linearity (\textbf{red}); in all other cases shear-hardening is absent and $\epsilon_{\rm Y}$ increases or remains nearly constant (\textbf{blue}).
    }
    \label{fig:simulation_yielding}
\end{figure*}

\begin{figure*}[ht]
    \centering
    \includegraphics[width=\linewidth]{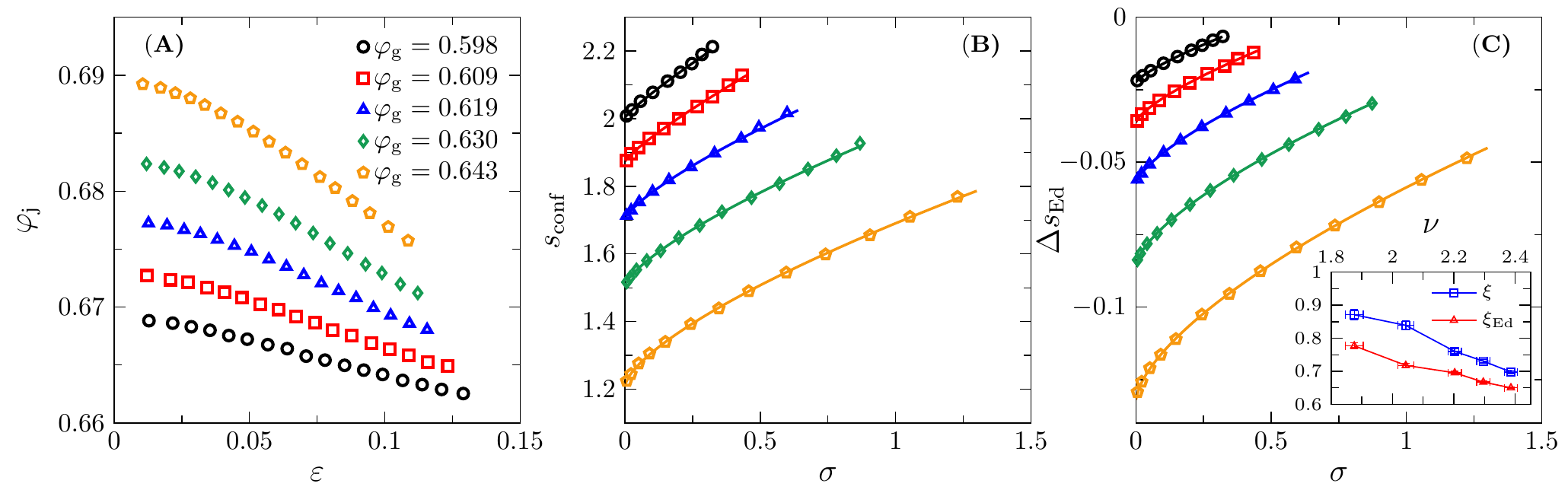}
    \caption{{\bf
    {Entropy evolution} of simulated granular matter under shear.} \textbf{(A)} Jamming density $\varphi_{\rm j}$ versus strain $\epsilon$, for a few different $\varphi_{\rm g}$.
    { \textbf{(B)} Configurational entropy $s_{\rm conf}$ from ~Ref.\cite{berthier2017configurational} and \textbf{(C)} Edwards entropy $\Delta s_{\rm Ed}$  as a  function of stress $\sigma$, fitted to Eq.~(\ref{eq:s0}) (solid lines, see SM Sec.~S6 for a discussion on the fitting). The fitting parameters ${\xi}$ and $\xi_{\rm Ed}$ as a function of the non-linear exponent $\nu$ are plotted in the inset of \textbf{(C)}.} The error bars represent the standard error of the fitting exponents.
    \label{fig:simulation_entropy}}
\end{figure*}

\subsection*{Simulations of a granular matter model}

We perform computer simulations of a 3D granular matter model, which consists of poly-disperse spheres interacting via frictionless short-range repulsive forces (see Materials and Methods).   The system is compression quenched from initial configurations that are equilibrated at $\varphi_{\rm g}$ using an efficient swap algorithm (see Materials and Methods),  to zero temperature where it jams randomly at $\varphi_{\rm j}$. Thus $\varphi_{\rm g}$  can be understood as a glass transition density, quantifying the degree of annealing (the larger $\varphi_{\rm g}$, the deeper annealing). Thanks to the swap algorithm, we are able to prepare ultra-stable states corresponding to deep annealing. This is the key reason to observe significantly stronger non-linear elasticity, compared to previous numerical studies~\cite{kawasaki2020shear}.
Additional data are provided in SM for 2D models {without (Sec.~S7) and with (Sec.~S8) friction}, and systems mechanically annealed by cyclic shear (Sec.~S6) , confirming that the reported behavior is qualitatively insensitive to friction, dimensionality and preparation protocols.

The above compression quenching procedure generates isotropic configurations at zero temperature and a finite pressure $P_{0}>0$, where $P_{0} = P(\epsilon =0)$ characterizes the distance to isotropic jamming, $P_{0} \sim \varphi_0 - \varphi_{\rm j}$. These isotropic configurations serve as unstrained ($\epsilon = 0$) reference states to athermal quasi-static shear (AQS) with simple strain deformations under constant volume conditions (see Materials and Methods). For a small $P_{0} = 10^{-2}$, the stress-strain curves in Fig.~\ref{fig:simulation_yielding}(A) display clear non-linear elasticity, following the scaling law Eq.~(\ref{eq:nonlinear_scaling1}). The exponent  $\nu > 1$ reveals shear-hardening behavior ($G$ increases with $\epsilon$), which is equivalent to a dilatancy effect under a constant pressure condition. The stress-strain curve is reversible before yielding in cyclic shear ({Fig.~S13 in SM}), confirming that the observed non-linearity has a dominating elastic origin.

The yield stress $\sigma_{\rm  Y}$ increases, and the yield strain $\epsilon_{\rm Y}$ decreases, with $\nu$ (see Fig.~\ref{fig:simulation_yielding}(B)), consistent qualitatively with our theoretical predictions in Fig.~\ref{fig:max}(B). Note that near jamming, the unstrained shear modulus $G_0 \approx 0$ and the unstrained bulk modulus $K_0$ is finite~\cite{o2003jamming}, thus we have set the Poisson's ratio $\mathfrak{r}$ close to one in the theory (see Fig.~\ref{fig:max}(B)). The negative correlation between the yielding strain $\epsilon_{\rm Y}$ and the degree of annealing $\varphi_{\rm g}$ is in contradiction with previous simulation~\cite{jin2017exploring, ozawa2018random}
and theoretical results~\cite{ozawa2018random, rainone2015following}, where $\epsilon_{\rm Y}$ increases or nearly unchanged with the degree of annealing (see Fig.~\ref{fig:simulation_yielding}(C)). Note that in all those previous cases, the non-linear shear hardening effect is absent. To confirm this point, additional simulations are performed at larger $P_{0}$, where shear hardening is compensated by strong plasticity, and $\epsilon_{\rm Y}$ correspondingly becomes either independent of $\varphi_{\rm g}$
{ (Fig.~S14)}
or slightly increasing with $\varphi_{\rm g}$
{ (Fig.~S16(B))}.
We thus conclude that the left-shifted yielding peak is caused by non-linear corrections to the elasticity, and therefore can not be captured by linear elasticity  theories~\cite{ozawa2018random}. Shear hardening disappears in  poorly annealed  (small $\varphi_{\rm g}$, see Fig.~\ref{fig:simulation_yielding}(A)) or over-compressed (large $P_{0}$, see~Fig.~S14 { and~S16(B)}) systems, which explains   why it was not observed in many previous simulations~\cite{o2003jamming, ozawa2018random}.
Deep annealing and isostaticity (the coordination number $Z=2d$) could be  two key ingredients to  this effect.

Next, we investigate the change of entropy during shear in simulations. Fig.~\ref{fig:simulation_entropy}(A) shows that the jamming density $\varphi_{\rm j}(\epsilon)$, at which $P$ vanishes upon decompression for the given $\epsilon$, decreases monotonically with $\epsilon$. According to the mean-field glass theory, $\varphi_{\rm j}$ is positively correlated with the configurational entropy $s_{\rm conf}$ of glass states~\cite{parisi2020theory}. Thus, Fig.~\ref{fig:simulation_entropy}(A) already suggests an increase of entropy induced by shear, consistent qualitatively with our theoretical prediction, Eq.~(\ref{eq:s0}).
Because the direct computation of the zero temperature entropy is
impractical for systems of thousands particles~\cite{martiniani2017numerical},
we estimate it indirectly by the following two independent ways. (i) In the first approach, we assume that the zero-temperature configurational entropy at $\varphi_{\rm j}$ is proportional to the finite-temperature configurational entropy $s_{\rm conf}(\varphi_{\rm g})$ of the corresponding parent liquid state at $\varphi_{\rm g}$.
Under this assumption, we
collect the data of $s_{\rm conf}(\varphi_{\rm g})$ for the same model from Ref.~\cite{berthier2017configurational} (estimated by taking the difference between the total entropy $s$ and the vibrational entropy $s_{\rm vib}$), and $\varphi_{\rm j}(\varphi_{\rm g})$ from Ref.~\cite{jin2021jamming}.
Combining  them with $\varphi_{\rm j}(\epsilon)$ in Fig.~\ref{fig:simulation_entropy}(A) and $\sigma(\epsilon)$ in Fig.~\ref{fig:simulation_yielding}(A) gives $s_{\rm conf}(\sigma)$ in Fig.~\ref{fig:simulation_entropy}(B).
{ The numerical value of $\xi$ is estimated  by fitting the data to Eq.~(\ref{eq:s0}).
 (ii) In the second approach, we consider the {\it Edwards entropy}, constructed using the framework of Edwards statistical mechanics of granular matter, which is a generalization of Boltzmann statistical mechanics to non-equilibrium, athermal systems~\cite{edwards1989theory,RevModPhys.90.015006}. The Edwards entropy is computed based on the fluctuations of the local Voronoi volumes~\cite{PhysRevLett.101.188001} (see SM Sec.~S6 for details), and has been applied to unstrained, isotropic granular systems, in both simulations~\cite{PhysRevLett.101.188001,PhysRevE.80.031301, jin2010first} and experiments~\cite{PhysRevLett.127.018002}.
Here, we apply the method to anisotropic systems
 under simple shear, using the Lees-Edwards boundary conditions~\cite{lees1972computer}.
The change of Edwards entropy $\Delta s_{\rm Ed}$ per particle under shear (where we have taken the jammed states at the lowest jamming density, or the J-point density~\cite{o2003jamming}, $\varphi_{\rm J} \approx 0.655$~\cite{jin2021jamming}, as the reference), is plotted in Fig.~\ref{fig:simulation_entropy}(C) and fitted to Eq.~(\ref{eq:s0}) to obtain $\xi_{\rm Ed}$.
The two approaches give close results on the exponents (see Fig.~\ref{fig:simulation_entropy}(C)-inset). The numerical values of $\xi$ and $\xi_{\rm Ed}$ are of the same order of the theoretical prediction  (Fig.~\ref{fig:max}(C)-inset), and decay  similarly with the non-linear exponent $\nu$.} We point out that the agreement remains to be mainly qualitative due to the phenomenological nature of our theories.

\section*{Discussion}

Our results shed lights on the correct and concrete physical interpretation of the theoretical models at hand~\cite{PhysRevD.102.069901, Baggioli:2020qdg}. In particular, the behavior of the entropy under shear suggests that the holographic models considered \cite{Baggioli:2020qdg} are phenomenologically closer to amorphous solids rather than crystalline systems.
On the other hand, the current
versions of theories do not incorporate marginal stability and isostaticity, and therefore cannot properly describe
the jamming transition. The theories assume that the system is always stable, instead of marginally stable, before the breaking point, and thus are applicable only to the regime away from jamming where the comparison to simulations is made. It would be extremely interesting to incorporate the necessary microscopic information, and extend our theoretical formulation towards the jamming transition to explore its critical properties.

Our simulations show that shear hardening universally exists in 2D/3D, frictionless/frinctional, and thermally/mechanically annealed granular models,  which suggests that the phenomenon could be directly relevant to a number of experimental systems (see Sec.~S9)~\cite{xing2021x, zhao2022ultrastable, wang2022experimental}.  In particular, ultra-stable shear jammed granular materials were realized in a recent experiment~\cite{zhao2022ultrastable}, making a direct test of the discussed correlations  possible in the  laboratory.

The approach presented here can be generalized to study more complex non-linear behaviors.
For example, very close to jamming, the stress-strain curve of granular matter typically displays  three consecutive elastic regimes with  shear~\cite{kawasaki2020shear}: linear ($\nu=1$), shear softening ($\nu<1$) and shear hardening ($\nu>1$). In this study, we mainly focus on the third regime (shear-hardening) at large strains.
We expect the entire stress-strain curve to be  captured by a generalized two-potential gravity theory. Further extensions can be made by taking into account the effects of finite temperature, finite shear rates and visco-elasticity, as well as non-equilibrium relaxational dynamics.

\section*{Materials and Methods}

\subsection*{Effective field theory}

Following Ref.~\cite{Nicolis:2015sra},
the EFT description is implemented in terms of  $d=2$ scalar fields ($I=1 \ldots d$), $   \phi^I(t,\textbf{x})\,=\,\langle \phi^I \rangle+\delta \phi^I(t,\textbf{x})$, which play the role of co-moving coordinates. The variations from their equilibrium positions $\delta \phi^I(t,\bf{x})$ coincide with the displacement fields used in standard elasticity theory. The EFT action is built only in terms of the derivatives of the fields, reflecting the invariance under the shift symmetry $\phi^I \rightarrow \phi^I+a^I$ with $a^I$ constants. This symmetry follows from identifying the effective fields with phonons, which can be understood as Goldstone bosons for the spontaneously broken translations~\cite{Leutwyler:1996er} -- fluctuations around the non-trivial vacuum expectation values $\langle \phi^I \rangle = x^I$, with zero energy.

The only independent scalar objects that can be constructed out of the derivatives of the displacement fields (in $d=2$) are,
$X\,\equiv\,\mathrm{Tr}\,\left[\partial_\mu \phi^I \partial^\mu \phi^J\right]$ and
$Z\,\equiv\,\mathrm{Det}\,\left[\partial_\mu \phi^I \partial^\mu \phi^J\right]$,
where the index $\mu=(t,\bf{x})$ collectively describes the set of spacetime coordinates. The scalar potential then becomes a generic function $\mathcal{V}\left(X,Z\right)$
and it is the only ingredient that must be provided in the theory. In most of our discussion, we will consider a power-law potential, which is reminiscent of the so-called hyper-elastic models, and takes the form,
$ \mathcal{V}\left(X,Z\right)\,=\,X^A\,Z^{(B-A)/2}$.
Here $A$ and $B$ are two phenomenological model parameters that cannot be fixed without first-principle calculations.
The stress tensor of the theory can be derived using the standard quantum field theory variational prescription and the dispersion relation of the low-energy excitations by computing the action for the fluctuations around equilibrium at second order. More technical details are presented in SM Secs.~S1 and S2.

\subsection*{Gravity theory}

The gravitational description takes advantage of the so-called holographic duality in which a gravity system in a ($d+2$)-dimensional spacetime is mapped to a many-body system ``defined" on its ($d+1$)-dimensional boundary. The boundary system is referred to as a ``hologram" of the bulk. Thanks to the duality, introducing dissipative mechanisms and the effects of temperature becomes easy. More precisely, we use the bottom-up holographic duality in the large $\mathcal{N}$ limit. Here, the parameter $\mathcal{N}$ is interpreted as the number of effective degrees of freedom in the dual field theory \cite{Ammon:2015wua}. Large $\mathcal{N}$ in the gravity description corresponds to $L \gg l_p$ with $L$ the AdS length-scale and $l_p$ the Planck scale.
This framework is more general than the original, string theory inspired, anti-de Sitter/conformal field theory (AdS/CFT) correspondence~\cite{Maldacena:1997re} and it applies also to systems which are not critical (i.e., without conformal invariance). Concretely, the infrared (IR) physics of our gravitational system is governed by a non-relativistic geometry with an AdS$_2$ fixed point which does not enjoy the conformal group. Interestingly, this type of geometry shares some features with amorphous systems and spin glasses, and in particular, has a finite entropy at zero temperature~\cite{PhysRevB.100.205108}. Finally, in the large $\mathcal{N}$ limit (or equivalently $L \gg l_p$), all quantum loops in the boundary field theory are suppressed by factors of $1/\mathcal{N}$ and the corresponding physics is effectively classical. All these assumptions justify the validity of our framework to describe the nonlinear rheology of classical particles in amorphous systems. The framework thus allows us to perform simple and robust computations of
{ important physical}
observables such as the entropy. Our computations are based on the non-linear generalization~\cite{PhysRevLett.114.251602,Alberte:2015isw} of the known holographic axion model~\cite{Baggioli:2021xuv}.

The benchmark model uses a bulk potential form, $W(\mathcal{X},\mathcal{Z})\,=\,\mathcal{X}^\mathfrak{a}\,\mathcal{Z}^{(\mathfrak{b}-\mathfrak{a})/2}$, where $\mathcal{X}=\frac{1}{2}\mathrm{Tr}\left[\mathcal{I}^{IJ}\right]$ and $\mathcal{Z}=\mathrm{Det}\left[\mathcal{I}^{IJ}\right]$ with $\mathcal{I}^{IJ}\,\equiv\,\partial_\mu \varphi^I \partial^\mu \varphi^J$. Importantly, the index $\mu$ here spans a 4-dimensional spacetime $(t,\textbf{x},u)$ with $\textbf{x}\equiv (x,y)$.
While the bulk potential $W(\mathcal{X},\mathcal{Z})$ has similar form as the one in EFT, the connection between the bulk potential and the dual EFT potential is very non-local and subtle. To avoid any clutter we will always use different symbols for bulk quantities and EFT ones. The stress tensor of the dual field theory can be extracted from the gravitational action by using the standard holographic dictionary while the entropy by utilizing the famous Bekenstein-Hawking Area law. See SM Secs.~S3-5 for more details.

\subsection*{Granular model of 3D frictionless soft spheres}

The model~{ \cite{BCNO2016PRL, berthier2016growing, jin2021jamming}} is composed of $N = 8000$ SSs, with a diameter distribution  $P(D) \sim D^{-3}$, where $D_{\rm min} \leq D \leq D_{\rm  min}/0.45$.
Two spheres interact via a potential $V(r_{kl}) = \frac{k_v}{2}(1 - \frac{r_{kl}}{D_{kl}})^2$, only if their separation $r_{kl}$ is less than their mean diameter $D_{kl} = (D_k + D_l)/2$; otherwise, $V=0$. The unit of length is the average diameter of all particles, the unit of energy is $10^3 \times k_v$, and all particles have the same unit mass. Simulation data  are averaged over 96 independent samples.

\subsection*{Swap algorithm}

The SS granular configurations are quenched from equilibrium states at $\varphi_{\rm g}$~\cite{berthier2016growing,jin2021jamming}. To prepare these equilibrium configurations, we use HS potential and a very efficient swap Monte Carlo (MC) algorithm~\cite{BCNO2016PRL}. The HS configurations are generated by the hybrid of two different kinds of moves, the standard moves and the swap MC moves of exchanging the diameters of two randomly picked spheres. The swap moves are accepted only if the resulting configuration does not violate the HS constraint. Such moves help particles breaking out of cages formed by their neighbors and diffusing freely, and hence, facilitate the equilibration procedure. With the aid of this algorithm, we obtain equilibrium HS configurations over a wide range of $\varphi_{\rm g}$.

\subsection*{Simulation protocol of compression quench}

Once an equilibrium HS configuration is obtained by the swap algorithm, we switch off the temperature and switch to the SS potential. The system is then quenched to jamming density by a series of athermal quasi-static compression and decompression~\cite{o2003jamming}. If the system is jammed, i.e., the energy per particle is larger than $10^{-13}$, the system is decompressed; otherwise compressed. During each  compression  (decompression) step, we instantaneously inflate (deflate) the spheres to increase (decrease) the packing density by $\delta \varphi$, and then minimize the total potential energy using the FIRE algorithm~\cite{bitzek2006structural}. The energy minimization stops when the averaged residual force per particle is less than $10^{-11}$, which means that the configuration has reached a mechanically stable state.
The initial $\delta \varphi = 10^{-4}$;
it is then reduced by a factor of two, whenever the state alters from jammed to unjammed (in the meanwhile, we switch from decompression to compression), or vice versa. This procedure stops when $\delta \varphi< 10^{-6}$ and the system is jammed. The jammed configurations are quasi-statically compressed by $\delta \varphi = 10^{-5}$ to the target pressure $P_{0}$ to obtain unstrained configurations at $\epsilon = 0$.

\subsection*{Simulation protocol of athermal quasi-static shear}

Shear is performed under the athermal quasi-static and Lees-Edwards boundary~\cite{lees1972computer} conditions. Each shear step ($\delta \epsilon = 10^{-4}$) involves an affine transformation of coordinates, and then an energy minimization using the FIRE algorithm~\cite{bitzek2006structural}. The energy minimization stops when the average residual force per particle is less than $10^{-11}$.
The pressure $P$ and stress $\sigma$ (shear is applied in the $xy$ plane) is calculated from the Virial formula,
\beq
P =  \frac{1}{3V} \sum_{\langle kl \rangle } \mathbf{r}_{kl} \cdot \mathbf{f}_{kl}\,,\quad \sigma=-\frac{1}{V}\sum_{\langle kl \rangle } r_{kl,x} f_{kl,y}\,,
\eeq
where $\mathbf{r}_{kl}$ and $ \mathbf{f}_{kl}$ are the center-to-center vector and force between particles $k$ and $l$ ($r_{kl,x}$ and $ f_{kl,y}$ are the  $x$ and $y$ components), $V$ is the volume of simulation box, and $\langle kl \rangle$ stands for all contacting pairs.

\section*{Aknowledgments}
We thank Oriol Pujolas and Alessio Zaccone for providing useful comments and discussions.

\noindent{\bf Funding:} M.B. acknowledges the support of the Shanghai Municipal Science and Technology Major Project (Grant No.2019SHZDZX01). L.L. acknowledges the supports in part by NSFC No.12122513, No.12075298 and No.11991052, and by the CAS Project for Young Scientists in Basic Research YSBR-006. Y.J. acknowledges funding from {Project 12161141007,} Project 11974361 and  Project 11935002  supported by NSFC, and the Key Research Program of Frontier  Sciences, Chinese Academy of Sciences, Grant NO. ZDBS-LY-7017.  The authors from ITP acknowledge funding from Project 12047503 supported by NSFC and the Key Research Program of the Chinese Academy of Sciences, Grant NO. XDPB15.
The simulations were performed using the HPC Cluster at ITP-CAS.

\noindent{\bf Author contributions:} T.J., L.L. and M.B. performed the theoretical computations.
D.P. and Y.J. performed the simulations. All the authors contributed to the writing of the manuscript and the discussion of the ideas behind it.

\noindent{\bf Competing interests:} Authors declare that they have no competing interests.

\noindent{\bf Data and materials availability:} All data are available in the main text or the supplementary materials.



\onecolumn
\renewcommand\thefigure{S\arabic{figure}}
\setcounter{figure}{0}
\renewcommand{\theequation}{S\arabic{equation}}
\setcounter{equation}{0}
\renewcommand{\thesubsection}{S\arabic{subsection}.}
\renewcommand\thesection{S\arabic{section}}

\centerline{\LARGE Supplementary Materials for}
\centerline{\LARGE Non-linear elasticity, yielding and entropy in amorphous solids}

\vskip10pt

\centerline{Deng Pan, Teng Ji, Matteo Baggioli, Li Li, Yuliang Jin}

\vskip25pt

This PDF file includes:
\begin{itemize}
\item Fig.~S1. Illustration of affine deformations.
\item Fig.~S2. Relation between the yielding point and the gradient instability.
\item Fig.~S3. Correlation between  the yielding strain and the non-linear elasticity exponent from the EFT.
\item Fig.~S4. Stress-strain and entropy-strain curves from the gravity theory.
\item Fig.~S5. Numerical check of the derivation of the entropy scaling in the gravity theory.
\item Fig.~S6. Low temperature entropy at finite shear strains in the gravity theory.
\item Fig.~S7. Power-law fitting of the shear hardening part on stress-strain curves.
\item Fig.~S8. Dependence of the yielding strain on the parameter $c$.
\item Fig.~S9. Distributions of projected contact angles under simple shear.
\item Fig.~S10. Single and average stress-strain curves.
\item Fig.~S11. Power-law fitting of the configurational entropy obtained in simulations.
\item Fig.~S12. Edwards entropy.
\item Fig.~S13. Reversibility test of shear hardening.
\item Fig.~S14. Stress-strain curves of over-compressed systems.
\item Fig.~S15. Stress-strain curve of mechanically trained systems.
\item Fig.~S16. Stress-strain curves of 2D systems.
\item Fig.~S17. Stress-strain curve of frictional systems.
\end{itemize}

\twocolumn
\section{Setup of the effective field theory}
The effective field theory (EFT) for solids, and elastic materials in general, is based on the construction of a zero temperature effective action in which the low-energy continuous degrees of freedom are the phonons -- the Goldstone modes of translational invariance \cite{Leutwyler:1996er}. From a more general perspective \cite{Nicolis:2015sra}, condensed matter or soft matter systems can be defined as low-energy phases
which break spontaneously the high-energy fundamental Poincar\'e group. As such, the various phases, i.e. solids, liquids, superfluids, etc., are in 1-to-1 correspondence with the different possible symmetry breaking patterns of the Poincar\'e group. For the case of solids, as we will see, our description is obviously not complete and in particular it neglects features such as plasticity, thermal effects, and the presence of an underlying lattice breaking rotational invariance { (in the case of crystalline solids)}. Importantly, this theoretical construction is not restricted to the linear elasticity regime but it can be easily extended toward the non-linear region. See \cite{doi:10.1142/S2010194513009744,Armas:2019sbe,Delacretaz:2017zxd,Baggioli:2021ntj} for a modern treatment of hydrodynamics and viscoelasticity specially in connection to the holographic framework.

Under these assumptions, the fundamental building blocks are given by a set of $d$ (number of spatial dimensions) single-valued scalar fields $\phi^I$ (implying no plasticity nor non-affine dynamics \cite{Baggioli:2021ntj,PhysRevLett.127.015501}), whose background solution is given by a coordinate dependent vacuum expectation value (VEV):
\begin{equation}
    \langle \phi^I \rangle = {\delta^I}_j\, x^j\,. \label{vevs}
\end{equation}
The solution above obviously breaks translational symmetry $x^I \rightarrow x^I + a^I$ spontaneously  since it selects a preferred reference frame which can be thought of as the equilibrium position of the atoms or molecules of the medium. The scalars $\phi^I$ serve as a set of co-moving coordinates. The VEVs in Eq.~\eqref{vevs} preserve the rotational invariance in the spatial plane; isotropy is a simplifying assumption which can be easily generalized.
While such an assumption might seem odd in the context of periodic crystalline structures, which are obviously incompatible with rotational symmetry at the microscopic level, it is definitely suitable for amorphous systems where no precise ordered lattice is present, specially at length scales larger with respect to the granularity of the system. A second and more fundamental constraint imposed is that of \textit{homogeneity} at large scales. We indeed assume that at large distances, or equivalently small momenta, the system looks homogeneous. From a technical point of view, this is equivalent to impose a global shift symmetry for the scalars $\phi^I$ which acts on them as $\phi^I \rightarrow \phi^I+a^I$, and it is in some sense reminiscent of their Goldstone nature (shift-symmetric fields).

Following on these lines, the fundamental tensorial object in the theory is the kinetic matrix:
\begin{equation}
    \mathcal{I}^{IJ}\,\equiv\,\partial_\mu \phi^I \partial^\mu \phi^J\,, \label{ma}
\end{equation}
where the Greek index $\mu$ runs on the spacetime coordinates $(t,\bf{x})$, while the Latin one only on the spatial subset. Notice that this object is Poincar\'e invariant, a necessary ingredient if we assume our ultraviolet (UV) theory to have such a symmetry. In $d$ dimensions, the scalar objects that can be constructed out of the matrix in Eq.~\eqref{ma} are given by the traces of its powers:
\begin{equation}
    X_{(n)}\,\equiv\,\mathrm{Tr}\left[\mathcal{I}^n\right]\,.
\end{equation}
It is customary and convenient to replace one of the traces with the determinant of the matrix $Z\equiv \mathrm{Det} \left[\mathcal{I}\right]$ using the Newton's identities. For example:
\begin{align*}
    &Z\,=\,\frac{1}{2}\,X_{(1)}^2\,-\,\frac{1}{2}\,X_{(2)} \qquad {\rm in}\qquad d=2\,;\\
    &Z\,=\, \frac{1}{6}\,X_{(1)}^3\,-\,\frac{1}{2}\,X_{(2)}\,X_{(1)}\,+\,\frac{1}{3}\,X_{(3)} \qquad {\rm in}\qquad d=3\,;\\
    &\dots\,.
\end{align*}
From now on, for simplicity, we will focus on a 2D system. There are no fundamental obstructions in generalizing it to arbitrary dimensions, but several computational complications appear and render the underlying physics blurred. The most general effective action for an isotropic and homogeneous system which breaks spontaneously translational invariance is given by
\begin{equation}
    S_{\rm EFT}\,=\,\int\,d^3x\,\mathcal{V}(X,Z)\,,\label{ac}
\end{equation}
where to avoid clutter we have defined $X\equiv X_{(1)}$. Notice that here the temperature is set to zero and no dissipative effects such as viscosities are considered. We reiterate that the fields $\phi^I$ are taken to be single valued and therefore no plastic effects are considered.
\begin{figure}[ht!]

    \includegraphics[width=0.98 \linewidth]{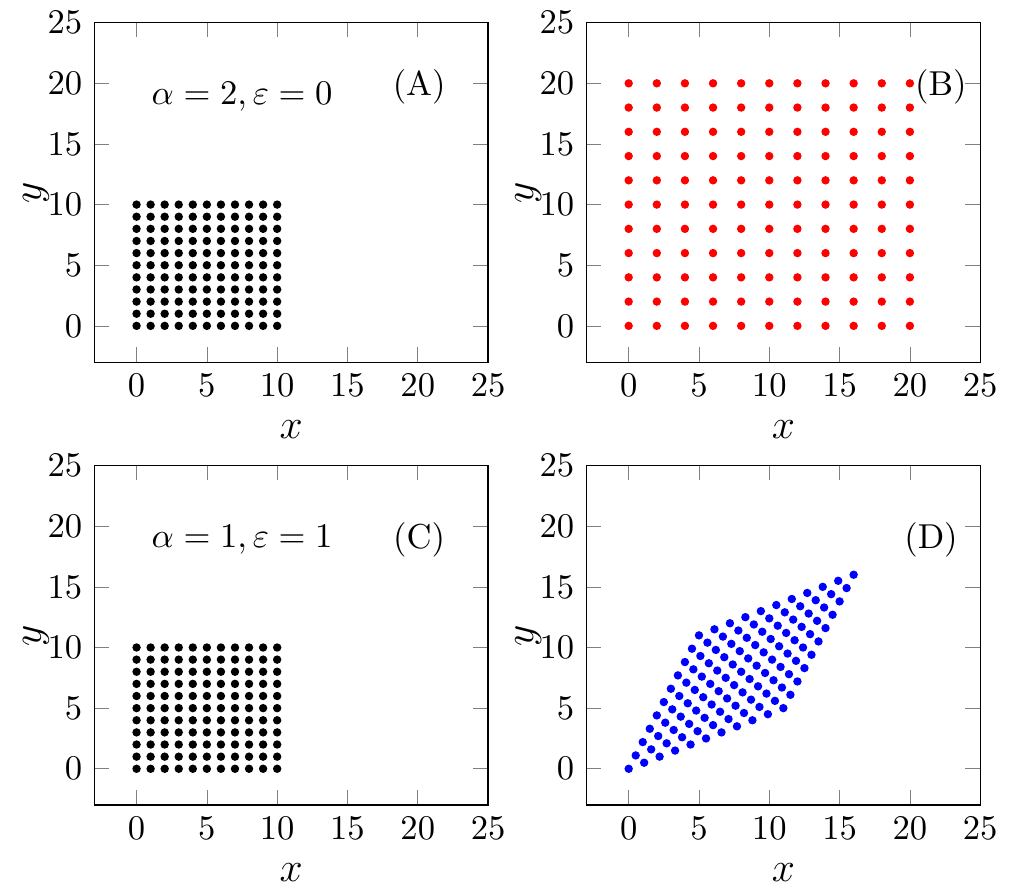}
    \caption{\textbf{Illustration of affine deformations}. \textbf{{(A) $\to$ (B):}} a pure bulk deformation corresponding to $\alpha=2,\varepsilon=0$ in Eq.~\eqref{finitedef}, changing the volume of the system but not the local angles. \textbf{{(C) $\to$ (D):}}
    a pure shear transformation corresponding to $\alpha=1,\varepsilon=1$ in Eq.~\eqref{finitedef}, modifying the angles but not the volume.}
    \label{app1}
\end{figure}

Importantly, the stress tensor of the system can be easily obtained as:
\begin{align}
T_{\mu\nu}\,=&\,-\,\frac{2}{\sqrt{-g}}\,\frac{\delta S_{\rm EFT}}{\delta g^{\mu\nu}}\,\Big|_{g=\eta}\nonumber\\&=\,-\,\eta_{\mu\nu}\,\mathcal{V}\,+\,2\,\partial_\mu \phi^I \partial_\nu \phi_I\,\mathcal{V}_X\, +\nonumber\\&+\,2\,\left(\partial_\mu \phi^I \partial_\nu \phi_I\,X\,-\,\partial_\mu \phi^I \partial_\nu \phi^J \,\mathcal{I}_{IJ}\right)\,\mathcal{V}_Z\,,
\end{align}
where we have finally taken the Minkowski background with the metric $g_{\mu\nu} =\eta_{\mu\nu}$.
For any time independent scalar field configurations, the stress-energy tensor components read
\begin{align}\label{rho}
&T^t_t\,\equiv\,\rho\,=\,\mathcal{V}\,,\\\label{pressure}
& T^x_x\,\equiv\,-\,p\,=\,\mathcal{V}\,-\,X\,\mathcal{V}_X\,-\,2\,Z\,\mathcal{V}_Z\,,\\\label{txy}
&T^x_{y}\,=\,2\,\partial_x \phi^I \partial_y \phi^I\,\mathcal{V}_X\,,
\end{align}
where $\mathcal{V}_X\equiv\partial \mathcal{V} /\partial X,$ etc. Here $\rho$ is the energy density and $p$ the mechanical pressure, and the position of the indices is irrelevant since we only consider a flat background metric $g_{\mu\nu}=\eta_{\mu\nu}$. To make a connection with the standard non-relativistic description of elasticity, $T_{ij}$ corresponds to the stress which is usually denoted as $\sigma_{ij}$ and from now on it will be denoted in that way.

At this point, the full dynamics of the low energy degrees of freedom, the phonons, can be obtained by considering small fluctuations around the equilibrium positions Eq.~\eqref{ma}:
\begin{equation}
    \phi^I\,=\,\langle \phi^I \rangle + \delta \phi^I\,,
\end{equation}
and by expanding the action~\eqref{ac} in terms of the new $\delta \phi^I$ fields. This will be sufficient to obtain the linear elastic response of the system but not the fully non-linear one. In order to go to the non-linear order, we need to generalize the scalar configuration~\eqref{vevs}. In particular, we consider the more generic Ansatz:
\begin{equation}\label{finitedef}
\phi_{\text{str}}^I=
O^I_Jx^J\;,\quad
O^I_J\,=\,\alpha\begin{pmatrix}
\sqrt{1+\varepsilon^2/4} & \varepsilon/2 \\
\varepsilon/2 & \sqrt{1+\varepsilon^2/4}
\end{pmatrix}\,.
\end{equation}
This solution for the scalars is not an equilibrium solution, but it describes a deformed configuration. In particular, the two parameters $\alpha$ and $\varepsilon$ are directly related to a background bulk strain and a background shear strain as shown in Fig.~\ref{app1}. The $\alpha$ parameter controls the change of volume in the system; on the contrary, the $\epsilon$ parameter implements a pure shear deformation where the corresponding strain tensor contribution reads:
\begin{equation}
    \varepsilon_{xy}\,=\,\varepsilon\,,
\end{equation}
therefore not modifying the volume of the system.
Using~\eqref{txy} combined with our scalar configuration~\eqref{finitedef}, one can obtain the fully non-linear stress-strain curve:
\begin{equation}\label{sigma}
\sigma(\varepsilon)\equiv T_{xy}=2\,\varepsilon\,\sqrt{1+\frac{\varepsilon^2}{4}} \;
\mathcal{V}_X\left(2+\varepsilon^2,1\right)\,.
\end{equation}

Linearizing the expression above and using the standard linear elasticity relation $\sigma_{xy}=G_0\varepsilon_{xy}+\dots$, we can determine the unstrained elastic shear modulus to be:
\begin{equation}\label{Glin}
G_0= 2\,\mathcal{V}_X(\bar{X},\bar{Z})\,
\end{equation}
where $\bar{X}=2,\bar{Z}=1$ are just the values at the equilibrium configuration Eq.~\eqref{vevs}. Here, the unstrained shear modulus $G_0$ is simply given by the zero strain value of the non-linear one $G\equiv d\sigma/d\epsilon$.
The same procedure can be followed to obtain all the linear elasticity properties of the systems such as the unstrained bulk modulus $K_0$, the Poisson ratio $\mathfrak{r}$ and many more in terms of the potential $\mathcal{V}$. To make an instructive analogy, the potential $\mathcal{V}$ plays exactly the same role of the strain energy function utilized in the standard treatment of non-linear elasticity~\cite{fu_ogden_2001,Beatty1996}.
Finally, the dispersion relation of the phonon modes on the deformed background Eq.~\eqref{finitedef} can be obtained by expanding the original action Eq.~\eqref{ac} up to quadratic order in the fluctuations $\delta \phi^I$. The details can be found in~\cite{PhysRevD.102.069901}. Importantly, because of the broken rotational invariance of the background Eq.~\eqref{finitedef}, transverse and longitudinal phonons are coupled together and the speed of propagation is a function not only of the parameters $\alpha,\varepsilon$ in Eq.~\eqref{finitedef} and the potential $\mathcal{V}$ but also of the propagation angle $\theta$ \cite{PhysRevD.102.069901}. This anisotropy is more and more evident for large values of the external shear strain and it can be consistently neglected at small (enough) strain.

\section{Breaking point from effective field theory}
Once the dynamics of the EFT is known (as explained in the previous section), one could ask whether and when the non-linear elastic response will display any sort of instability corresponding to \textit{breaking points} $( \varepsilon_{\rm Y}, \sigma_{\rm Y})$. In a non-relativistic system, one should not worry about any superluminality or causality issue. Therefore, the first and most dangerous instability is the so-called \textit{gradient instability}, which corresponds to the point at which the (strain dependent) velocity of sound $v_s$ becomes imaginary and therefore the system dynamically unstable~\cite{PhysRevD.102.069901}. More precisely, stability requires that:
\begin{equation}
    v_s^2(\varepsilon,\alpha,\theta)\,\geq\,0\,,
\end{equation}
and it imposes strong constraints on the maximum strain that the system can support:
\begin{equation}
    v_s^2( \varepsilon_{\rm Y})\,=\,0\,.
    \label{eq:yielding_def}
\end{equation}

At least qualitatively, this point of instability can be associated with the yielding point in the non-linear stress-strain response and with the breakdown of the elastic response. Let us consider, for example, a simple linear isotropic system where the propagation speed of  the transverse phonons is given as usual by $v_T^2=G_0/\rho_m$ where $\rho_m$ is the mass density and $G_0$ the unstrained shear elastic modulus, which can be extracted from $\sigma= G_0 \varepsilon + \mathcal{O}\left(\varepsilon^2\right)$. By promoting this relation to non-linear level, $\sigma(\varepsilon)$, one could define a strain dependent velocity  $v_T^2(\varepsilon)$ and a strain dependent non-linear elastic modulus $G(\varepsilon)\equiv d\sigma/d\varepsilon$ as follows\footnote{It is not guaranteed that the nonlinear strain dependent velocity will still take the form $v_T^2=G/\rho_m$ where $G_0$ is simply replaced with the strain dependent non-linear shear modulus $G(\varepsilon)$. Indeed, our computations~\cite{PhysRevD.102.069901} show that it is not the case. Nevertheless, this approximation is sufficient to provide an intuitive relation between the yielding point and the gradient instability.}:
\begin{equation}
    v_T^2(\varepsilon)=G(\varepsilon)/\rho_m\,.
    \label{eq:vT}
\end{equation}
Continuing on these lines, one could derive the maximum strain as the point at which:
\begin{equation}
     v_T^2( \varepsilon_{\rm Y})\,=\,0\qquad \longrightarrow \qquad G( \varepsilon_{\rm Y})\,=\,0\,.
\end{equation}

\begin{figure}[ht!]
    \centering
    \includegraphics[width= \linewidth]{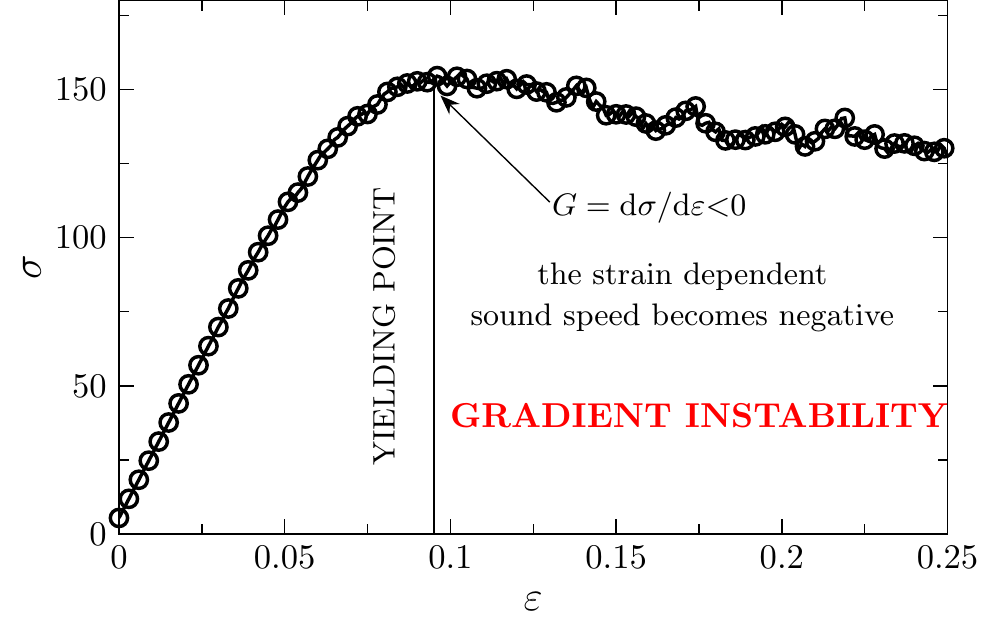}
    \caption{\textbf{Relation between the yielding point and the gradient instability.} An illustration of the relation between the maximum strain-stress located at the yielding point and the gradient instability signalled by the strain dependent sound speed becoming negative. This picture is intuitive and it assumes that the strain dependent velocity is given by $v^2(\varepsilon)=G(\varepsilon)/\rho$ which is only approximately true. The data are from the simulations of \cite{Baggioli:2021iom}.}
    \label{insta}
\end{figure}

By using the definition of the strain dependent elastic modulus as $G(\varepsilon)\equiv d\sigma/d\varepsilon$, one can identify the maximum strain and the instability point with the condition:
\begin{equation}
    \frac{d \sigma}{d \varepsilon}\,\Big|_{ \varepsilon_{\rm Y}}\,=\,0\,,
    \label{eq:yielding_def1}
\end{equation}
which coincides exactly with the yielding point in the non-linear stress strain curve (see Fig.~\ref{insta}).

Note that the above discussion is rather intuitive since Eq.~(\ref{eq:vT}) is valid only in the linear regime and it indeed receives corrections at the non-linear level. This implies that the two conditions Eqs.~\eqref{eq:yielding_def} and \eqref{eq:yielding_def1} are not equivalent for systems with non-linear elasticity. Because of the absence of plasticity (which results in a monotonically increasing stress-strain curve), in the theory, we will always use the precise criterion given in Eq.~\eqref{eq:yielding_def}.
On the other hand, to define the breaking point (yielding point) in simulations, we will use for simplicity the  criterion in Eq.~\eqref{eq:yielding_def1} (see Sec.~S6 for details). This is because that in simulations, only for unaveraged single-sample stress-strain curves, yielding can be defined by a sharp stress drop, which is the signal of a global instability and in this sense closer to the criterion defined in Eq.~\eqref{eq:yielding_def}. Once the average over samples is taken (which is usually necessary  in simulations), the average stress-strain curve becomes smooth and the condition Eq.~\eqref{eq:yielding_def1} applies.
Nevertheless, as shown explicitly using the numerical simulations in Sec.~S6, the two conditions give compatible results: the average yielding point defined based on Eq.~\eqref{eq:yielding_def1} is close to the individual breaking points of each sample.
Therefore, at least from a  qualitative point of view, the two criteria in Eqs.~\eqref{eq:yielding_def} and \eqref{eq:yielding_def1} are interchangeable in simulations.

In order to make concrete predictions, we make use of a non-relativistic potential:
\begin{equation}
    \mathcal{V}\,=\,\rho_0\,\left(\sqrt{Z}+ { \mathcal{V}_0}^2\,\left(\frac{X}{2}\right)^A\,Z^{(B-A)/2}\right)\,, \label{ww}
\end{equation}
as done in Section IV of~\cite{PhysRevD.102.069901}. Here, ${ \mathcal{V}_0} \ll 1$ is a small parameter that controls the value of the sound speed. Importantly, the first new term in~\eqref{ww} does not affect the dynamics of the stress-strain curve but only the size of the energy density $\rho$. Concretely, this is a sensible way of taking the non-relativistic limit for a potential of the type $ \mathcal{V}\,\sim X^A\,Z^{(B-A)/2}$ and allow for a speed of sound which is much smaller than the speed of light $c$, as in realistic systems. The most important effect of this non-relativistic limit is that of enlarging the allowed values of the parameters $A,B$ to the range $A>0,B>1$, and avoiding the previous limitation coming from relativistic invariance. We do not repeat the computations of~\cite{PhysRevD.102.069901} here but only state and discuss the main results related to the non-relativistic potential Eq.~\eqref{ww}.

\begin{figure}[ht!]
    \centering
    \includegraphics[width= \linewidth]{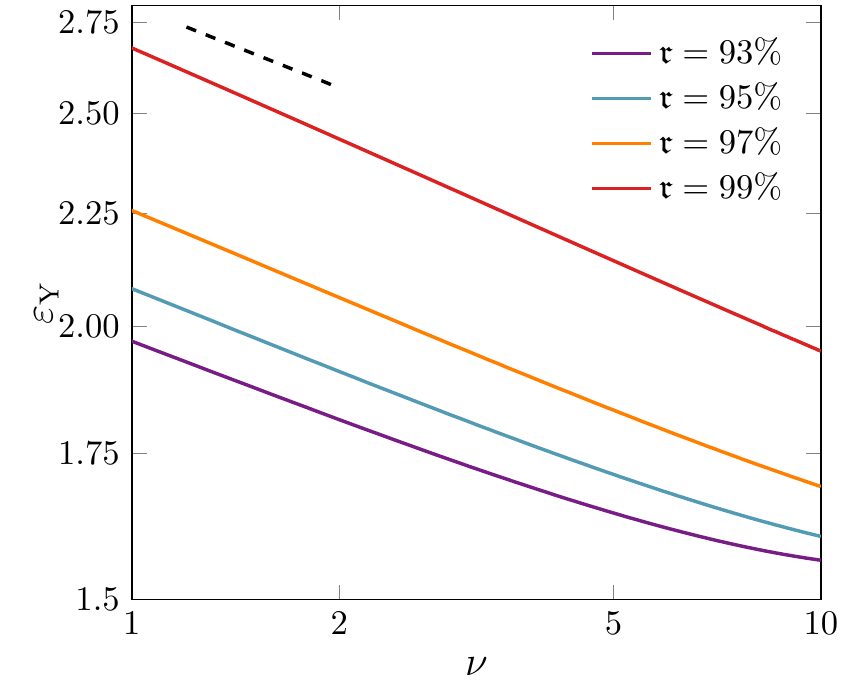}
    \caption{\textbf{Correlation between  the yielding strain and the non-linear elasticity exponent from the EFT.} The yielding point strain $\varepsilon_{\rm Y}$ in function of the non-linear elasticity exponent $\nu$ for different values of the Poisson ratio $\mathfrak{r}$ close to the incompressible limit $\mathfrak{r}=1$. The dashed line indicates the power law scaling $\nu^{-1/8}$ of Eq.~\eqref{tt2}.}
    \label{cc}
\end{figure}

For this specific potential, by neglecting the effect of anisotropy (see~\cite{PhysRevD.102.069901}), the strain dependent speed of sound is given by:
\begin{align}
  & v_s^2(\varepsilon)= 4^{-A} A\, v^2 \left(\varepsilon ^2+2\right)^{2 A-1} \nonumber\\&\left(A^2 \left(\varepsilon ^2+4\right) \varepsilon ^2-A \,B
   \left(\varepsilon ^2+4\right) \varepsilon ^2+4\, A+4 (B-1) B\right)\,.
\end{align}
By searching for the roots of the above expression, we can identify the maximum strain
\begin{equation}
    \epsilon^2_{\rm Y}\,=\,2 \sqrt{\frac{A (B-A)+A+(B-1) B}{A (B-A)}}-2\,,
    \label{eq:epsilon_max_EFT}
\end{equation}
and the breaking stress
\begin{equation}
    \sigma_{\rm Y}\,=\, A \,\epsilon_{\rm Y} \, \sqrt{\epsilon^2_{\rm Y}+4}\, \left(\epsilon^2_{\rm Y}+2\right)^{A-1}\,,
    \label{eq:sigma_max_EFT}
\end{equation}
which can be easily re-written in terms of the non-linear exponent $\nu$ and the Poisson ratio $\mathfrak{r}$ using the definitions presented in the main text. More explicitly, Eq.~\eqref{eq:epsilon_max_EFT} becomes
\begin{align}
   \epsilon^2_{\rm Y}= 2 \sqrt{\frac{-\nu +M+(\nu -1) \mathfrak{r}+5}{-\nu +M+(\nu -1) \mathfrak{r}+1}}-2\,,
\end{align}
with $M=\sqrt{2 \nu -2 \nu  \mathfrak{r}^2+\mathfrak{r}^2-2 \mathfrak{r}+1}$. Expanding $\epsilon_{\rm Y}$ near the incompressible limit, $\mathfrak{r}=1$, we obtain
\begin{align}
    \epsilon_{\rm Y}=\frac{2^{3/4}}{(1-\mathfrak{r})^{1/8}}\nu^{-1/8}+\mathcal{O}(1-\mathfrak{r})^{1/8}\,.\label{tt2}
\end{align}
This result suggests that, in the incompressible limit (for large values of the Poisson ratio $\mathfrak{r}$), the maximum strain follows a power law in terms of the non-linear exponent $\nu$ given by
\beq
\epsilon_{\rm Y}\sim\nu^{-\kappa}
\eeq
where $\kappa= 1/8$. This analytical scaling is confirmed in Fig.~\ref{cc}.

Let us conclude this section with a few comments about these results. Eq.~\eqref{eq:epsilon_max_EFT} is a concrete prediction of the EFT formalism. It has to be considered as a qualitative but sharp correlation between the location of the breaking point (which as discussed above can be roughly identified with the yielding point) and the non-linear elastic properties of the material. In particular, the direct prediction of the EFT is that: \textit{the yielding strain diminishes with increasing the non-linear elastic exponent $\nu$}. Despite we are not able to formally and mathematically prove this statement for a generic potential $\mathcal{V}$, we have substantial evidences that this behaviour, at least from a qualitative point of view, is universal. This behaviour is reversed at very large values of the non-linear exponent $\nu$ (e.g. $\nu \approx 40$ for $\mathfrak{r}=97\%$) before which nevertheless the approximation scheme of the EFT breaks down and therefore its results cannot be trusted anymore. Moreover, the power-law behaviour in Eq.~\eqref{tt2} is a consequence of the precise power-law form of the EFT potential. In other words, despite the concrete number (in this case $-1/8$) is not universal, a power-law form of the stress-strain curve seems to always imply a power-law correlation between the maximum strain and the non-linear parameter $\nu$. Unfortunately, our data from the simulations are not enough to prove or disprove a possible power-law correlation between these two quantities. Finally, as discussed in~\cite{Baggioli2020}, the same qualitative behaviour is observed also in the gravitational model that we will describe in the next section.

\section{Setup of the holographic gravitational model}
The gravitational model makes use of the so-called gauge-gravity duality (or holographic) formalism~\cite{Baggioli:2019rrs} and it is deeply inspired by the EFT constructions~\cite{Nicolis:2015sra} just mentioned. The idea, introduced in~\cite{PhysRevLett.114.251602} and developed in~\cite{Alberte:2015isw} (see also~\cite{Baggioli:2021xuv} for a recent review), is to embed the EFT structure in Eq.~\eqref{ac} into an asymptotically anti-de Sitter curved spacetime by using a 4-dimensional bulk action:
\begin{equation}
    S\,=\,\int\,d^4x\,\sqrt{-g}\,\left[R\,-\,2\,\Lambda\,-\,2\,m^2\,W(\mathcal{X},\mathcal{Z})\right]\,,
\end{equation}
where $g$ is the determinant of the curved metric $g_{\mu\nu}$, $R$ the corresponding Ricci scalar, $\Lambda<0$ the negative cosmological constant and $m$ a dimensionful parameter which will be related to the graviton mass. There are two massless scalar bulk fields $\varphi^I$ with
$\mathcal{I}^{IJ}\,\equiv\,\partial_\mu \varphi^I \partial^\mu \varphi^J$ and $\mathcal{X}=\frac{1}{2}\mathrm{Tr}\left[\mathcal{I}^{IJ}\right]$ and $\mathcal{Z}=\mathrm{Det}\left[\mathcal{I}^{IJ}\right]$.

These models are usually called "homogeneous holographic models" and they exhibit several interesting features such as viscoelastic properties and propagating phonon modes~\cite{Baggioli2020,PhysRevLett.120.171602,Baggioli:2018bfa,Andrade:2019zey,Baggioli:2019elg,Ammon:2019apj,Baggioli:2019abx,PhysRevLett.114.251602,Alberte:2015isw,Baggioli:2021xuv,PhysRevResearch.2.022022,Alberte:2016xja}. It is important to stress that, a gravitational model with potential $W(\mathcal{X},\mathcal{Z})$ is not the dual of an EFT defined by the action Eq.~\eqref{ac} with the same potential. In other words, the connection between the bulk potential and the dual EFT potential is very non-local and subtle. To avoid any clutter we will always use different symbols for bulk quantities and EFT ones. Additionally, the idea that these gravitational models are the exact duals of the EFTs in \cite{Nicolis:2015sra} is not totally correct since the global symmetries of the EFTs are kept global in the bulk picture. This caveat is emphasized in \cite{Esposito:2017qpj}.

The massless scalar fields $\varphi^I$ are now living in the four dimensional curved bulk spacetime whose metric is taken to be:
\begin{align}\label{metric}
d s^{2}=\frac{1}{u^{2}}\left(-f(u) e^{-\chi(u)} d t^{2}+\frac{d u^{2}}{f(u)}+\gamma_{i j}(u) d x^{i} d x^{j}\right)\,,
\end{align}
where $u$ is the radial extra dimension spanning from the UV conformal boundary $u=0$ to the black hole horizon $u=u_h$ at which $f(u)$ vanishes. The spatial metric $\gamma_{ij}$ is generally not invariant under SO(2) rotations because of the background mechanical deformations. The scalar bulk fields $\varphi^I$ are dual, in the holographic sense, to the scalar operators $\phi^I$ used in the EFT description in previous sections and their background solution is taken as,
\begin{equation}\label{scal}
\begin{pmatrix}
\varphi^x \\
\varphi^y
\end{pmatrix}\,=
\alpha
\,\begin{pmatrix}
\cosh\left(\Omega/2\right) &  \sinh\left(\Omega/2\right) \\
 \sinh\left(\Omega/2\right) &  \cosh\left(\Omega/2\right)
\end{pmatrix}\,\begin{pmatrix}
x \\
y
\end{pmatrix}\,,
\end{equation}
in complete analogy with the EFT description in Eq.~\eqref{finitedef}. By comparing the two parametrizations, it is easy to identify the shear strain with:
\begin{equation}
    \varepsilon\,=\,2\,\sinh\left(\Omega/2\right)\,,
\end{equation}
while $\alpha$ corresponds to the bulk deformation.
Notice that the strain $\varepsilon$ is here taken as a background parameter, not as an infinitesimal deformation. This is the crucial technical point which allows us to compute the elastic response at a fully non-linear level.

Before proceeding, a very important observation has to be made. In the EFT description, the scalars solution in Eq.~\eqref{finitedef} breaks translational invariance spontaneously. In the gravitational setup, the picture is more complicated and it crucially depends on the asymptotics of the scalar fields $\varphi^I$ close to the boundary. For simplicity, let us consider a potential whose asymptotic boundary expansion takes the form:
\begin{equation}
    W(\mathcal{X},\mathcal{Z})\,=\,\mathcal{X}^\mathfrak{a}\,\mathcal{Z}^{(\mathfrak{b}-\mathfrak{a})/2}\,+\,\dots\,,
\end{equation}
where the ellipsis indicates subleading term with faster fall-off. Under this assumption, the asymptotic behaviour of the bulk scalars is given by:
\begin{equation}\label{exp}
   \varphi^I\,=\, \varphi_0^I\,\left(1\,+\,\dots\right)\,+\,\varphi_1^I\,u^{5-2\mathfrak{a}}\,\left(1\,+\,\dots\right),\quad \text{for} \quad u\rightarrow 0\,.
\end{equation}
Using the so-called standard quantization scheme rooted in the holographic dictionary \cite{Witten:1998qj}, the leading term in the above expansion must be identified with the external source for the dual field theory operator while the subleading term with its expectation value. The configuration in Eq.~\eqref{scal} is independent of the radial coordinate $u$ and it has to be identified with the term $\varphi_0^I$ in the expansion Eq.~\eqref{exp}. Following this logic, the solution in Eq.~\eqref{scal} is a VEV for the dual fields in the field theory side, as in Eq.~\eqref{vevs}, only if $5-2\mathfrak{a}<0$. Therefore, only in that regime, the gravitational theory describes the same physical systems as in the EFT presented in the previous sections. Indeed, the condition above ensures that in the holographic picture the translational invariance of the boundary field theory is broken spontaneously. This guarantees that the dual systems are solids with propagating phonons and well-defined elastic properties as desired. For more details about this point we refer to the vast literature on the topic~\cite{Alberte:2017cch,PhysRevLett.120.171602,Andrade:2019zey,Ammon:2019wci,Ammon:2019apj,Baggioli:2019abx,Baggioli:2019elg,Baggioli:2019mck,Ammon:2020xyv,Baggioli:2020edn}.

To continue, we define the spatial matrix
\begin{equation}
\gamma_{ij}(u)\,=\,\begin{pmatrix}
e^{a(u)}\,\cosh c(u) & \sinh c(u) \\
\sinh c(u) & e^{-a(u)}\,\cosh c(u)
\end{pmatrix}\,,
\end{equation}
which contains two independent functions $c(u),a(u)$ corresponding to the two different polarizations of the graviton. It is easy to show that one can set consistently $a(u)=0$ reducing this matrix to a single radial dependent function. Having defined our Ansatz, the background equations of motion read:
\begin{equation}\label{eqs}
\begin{split}
& f\, \left(u^2\, {c'}^2+12\right)+4 \left(m^2\,W-u \,f'-3 \right)\,=\,0\,,\\
&c''+c' \left(\frac{f'}{f}-\frac{2}{u}\right)-\frac{1}{4} u
   \,{c'}^3-\frac{2 \alpha ^2 \,m^2 \sinh (\Omega-c)\, W_\mathcal{X}}{f}=0\,, \\
 &  \chi '\,=\frac{1}{2}\,u \,{c'}^2\,,
\end{split}
\end{equation}
where we have taken $\Lambda=-3$ and the potential $W$ is evaluated on the background values $\bar{\mathcal{X}}\equiv \alpha ^2 u^2 \cosh (\Omega-c)$ and $\bar{\mathcal{Z}}\equiv \alpha ^4 u^4$.

Notice that the $\chi$ function is completely slaved to the $c$ function and therefore the final set of variables can be simply thought as the pair $(f,c)$. The system of~\eqref{eqs} can be solved numerically by imposing the presence of an event horizon at the location $u=u_h$ where $f(u_h)=0$ and $c(u_h)=c_h$. The UV boundary condition $\chi(u=0)=0$ is used to fix the time scale of the time coordinate so that the temperature at the UV boundary is equal to the standard Hawking temperature. Close to the UV boundary $u=0$, we have $f(u)=1$ and $\chi(u)=0$ together with the asymptotic expansion:
\begin{equation}
c(u)\,=\,\mathcal{C}_0\,\left(1\,+\,\dots\right)\,+\,\mathcal{C}_3\,u^3\,+\,\dots\,, \label{exp2}
\end{equation}
where, within the standard quantization scheme, $\mathcal{C}_0$ is identified with the source for the $T_{xy}$ operator and $\mathcal{C}_3$ with its expectation value $ \langle T_{xy} \rangle$. Finally, we will consider only setups with zero source for the stress tensor operator, $\mathcal{C}_0=0$. The equations of motion~\eqref{eqs} allow the scaling symmetry
\begin{equation}\label{symmetry}
u \rightarrow\lambda u,\quad \alpha\rightarrow \frac{1}{\lambda}\alpha,\quad (f, c)\rightarrow(f,c)\,,
\end{equation}
with $\lambda$ a constant. This is a consequence of scale invariance of the model. One can fix the symmetry defined in Eq.~\eqref{symmetry} by setting $\alpha=1$. This corresponds to considering only pure shear deformations which are volume-preserving.

The full non-linear elastic response can be studied by looking at the expectation value of the stress tensor operator $ \langle T_{xy} \rangle\equiv\sigma=\frac{3}{2}\mathcal{C}_3$ in function of the external mechanical strain  $\varepsilon\,=\,2\,\sinh\left(\Omega/2\right)$. Moreover, one could extend the analysis to the out-of-equilibrium time-dependent dynamics as initiated in \cite{Baggioli:2019mck}. In this work, we restrict our analysis to the shear sector. The interested reader can find the analysis of the longitudinal sector in \cite{Baggioli2020}. Finally, even in this model the maximally allowed strain and stress  can be obtained by looking at the gradient instabilities of the gravitational modes. A partial analysis of this sort has been presented in \cite{Baggioli2020}.

Before concluding this short summary of the gravitational model,
let us stress the benefits of using this description. (i) This dual formulation permits to introduce a finite temperature $T\neq 0$ without major difficulties. This would be impossible from the standard EFT point of view, in which dissipative and finite temperature effects are notoriously hard to consider~\cite{Endlich:2012vt,Liu:2018kfw}. (ii) In this scenario, the computation of the entropy of the system is particularly simple and it boils down to the estimation of the black hole entropy given by the famous Bekenstein-Hawking Area law~\cite{hawking1971gravitational,Bekenstein1972}.

\section{Scalings from the gravitational model}
Let us consider the gravitational model introduced in the previous section with a bulk potential of the type:
\begin{equation}
    W(\mathcal{X},\mathcal{Z})\,=\,\mathcal{X}^\mathfrak{a}\,\mathcal{Z}^{(\mathfrak{b}-\mathfrak{a})/2}\,,\label{sim}
\end{equation}
where the parameters $\mathfrak{a},\mathfrak{b}$ need to obey the following constraints:
\begin{align}
    &\mathfrak{a}\,\geq\,0\,,\,\mathfrak{b}\,\geq\,1,\qquad \qquad \text{no instabilities \cite{PhysRevD.102.069901}}\,,\\
    & \mathfrak{b}\,>3/2,\,\qquad \qquad \hspace{0.71cm} \text{positive shear modulus \cite{PhysRevLett.120.171602}}\,,\\
     & \mathfrak{b}\,>5/2,\,\qquad \qquad \hspace{0.71cm} \text{massless phonons \cite{PhysRevLett.120.171602}}\,.
\end{align}
Here, we assume always standard quantization for the bulk axion fields $\varphi^I$. Importantly, the duality between the EFT presented in the previous sections and the gravitational model is not simply given by the linear map $(A,B)\rightarrow (\mathfrak{a},\mathfrak{b})$.

\begin{figure}[ht!]
\begin{flushright}
    \includegraphics[width=\linewidth]{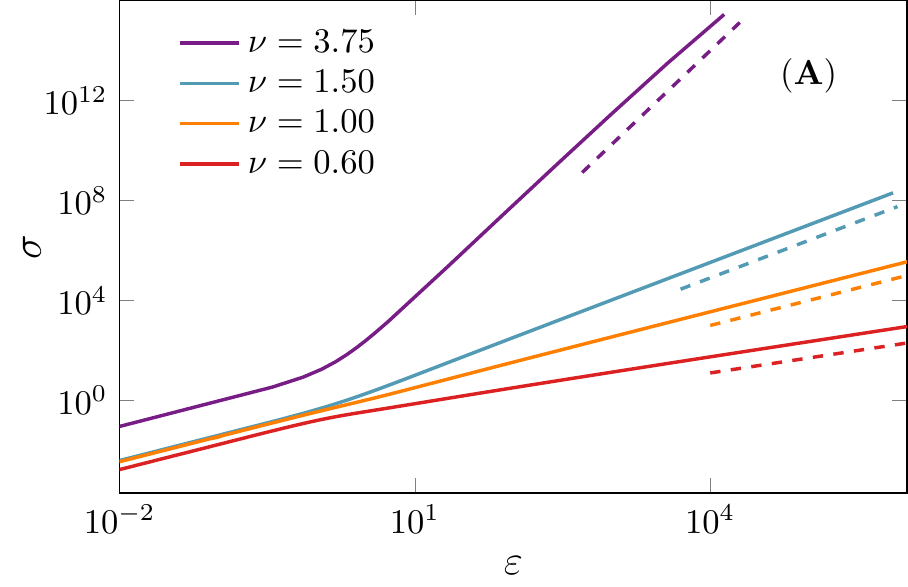}

    \vspace{0.5cm}

    \includegraphics[width=\linewidth]{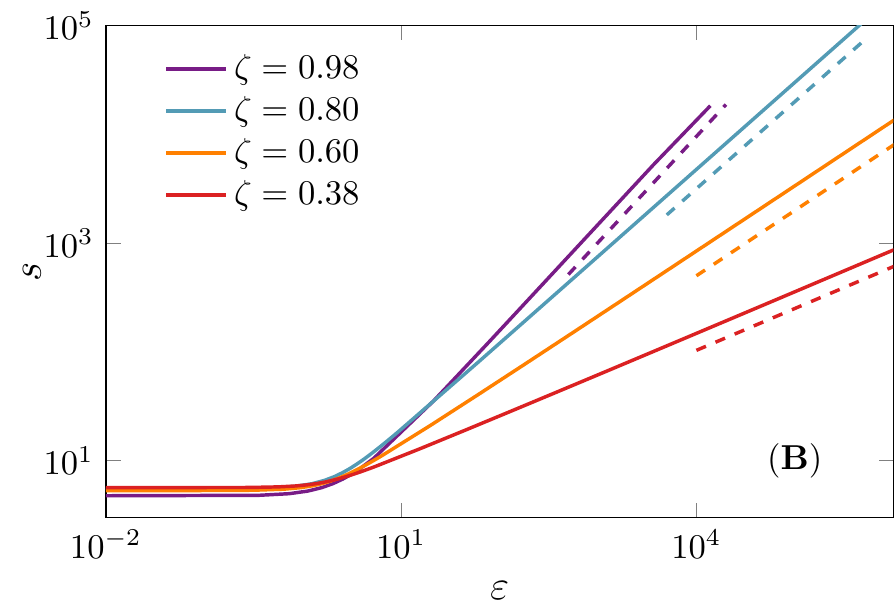}
\end{flushright}
    \caption{\textbf{Stress-strain and entropy-strain curves from the gravity theory.} \textbf{(A)} Non-linear stress strain curve for different choices of $\nu$. \textbf{(B)} Entropy density as a function of the background strain for different choices of $\zeta$. The dashed lines indicate the scalings in Eqs.~\eqref{sscal} and \eqref{escal}. The temperature is fixed to $T/m=0.1$.}
    \label{nonl}
\end{figure}

The full non-linear stress-strain curve can be extracted numerically for arbitrary values of $\mathfrak{a},\mathfrak{b}$ and analytically in the large temperature limit $T/m \gg 1$ \cite{Baggioli2020}. Some benchmark curves are shown in Fig.~\ref{nonl}.
\\

At large enough temperatures, after the linear elastic regime $\sigma= G_0 \varepsilon$, two different scaling regimes appear whose powers are given by:
\begin{equation}
    \nu_0\,=\,2\,\mathfrak{a}\,,\qquad \nu\,=\,\frac{3\,\mathfrak{a}}{\mathfrak{b}}\,.\label{sscal}
\end{equation}
Going towards low temperature, the first non-linear scaling $\nu_0$ disappears and the second one remains as the dominant one. These scalings can be analytically derived using the properties of the background strained geometry (see~\cite{Baggioli2020} for the detailed derivation). In summary, combining analytical and numerical methods we can predict and observe a non-linear stress-strain scaling law of the form:
\begin{equation}
\sigma(\epsilon)\,\sim \,\epsilon^{\nu}\,,\qquad \nu\,=\,\frac{3\,\mathfrak{a}}{\mathfrak{b}}\,.\label{scalfin}
\end{equation}

\begin{figure*}[ht!]
    \centering
  \includegraphics[width=0.3\linewidth]{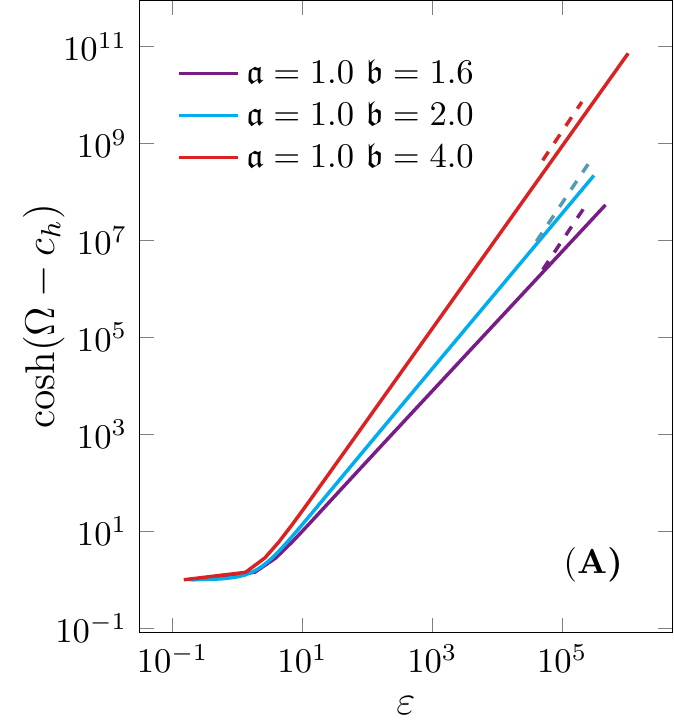} \qquad\quad
  \includegraphics[width=0.3\linewidth]{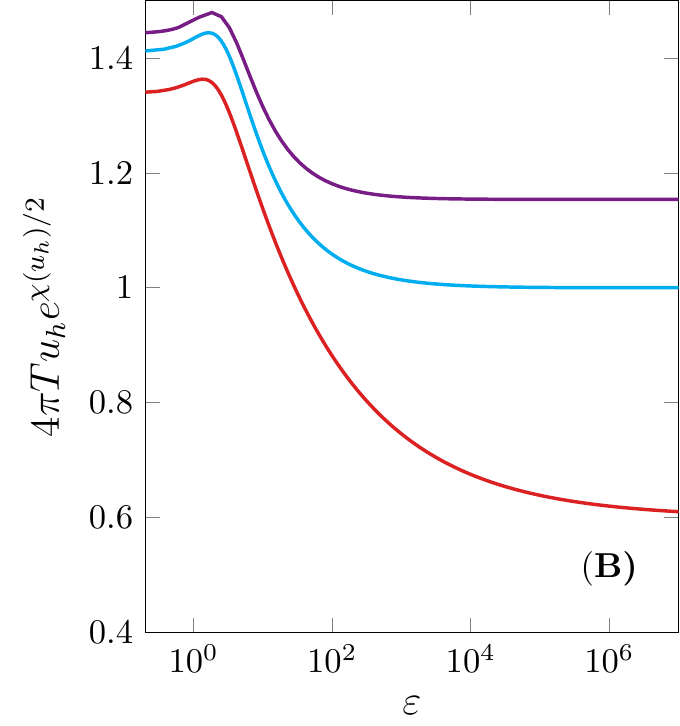}
    \caption{\textbf{Numerical check of the derivation of the entropy scaling in the gravity theory.} \textbf{(A)} The behaviour of the  { term $\cosh{(\Omega-c_h)}$} in Eq.~\eqref{c1} as the strain $\varepsilon$ increases for various values of $(\mathfrak{a},\mathfrak{b})$.
    The dashed lines indicate the nonlinear scaling $\varepsilon ^2 $. \textbf{(B)} The relation between $4 \pi u_h\,T\,e^{\chi(u_h)/2}$ and the strain $\varepsilon$ for various values of  $(\mathfrak{a},\mathfrak{b})$.  We have fixed $T/m=0.1$.}.
    \label{checks}
\end{figure*}

As a second step, we can compute the thermodynamic entropy $s=4\pi/u_h^2$ in function of the background strain. Our numerical results shown in Fig.~\ref{nonl} present a universal scaling law of the type:
\begin{equation}
\label{escal}
 s(\epsilon)\,\sim\,\epsilon^{\zeta}\,,\qquad \zeta\,=\,\frac{2\mathfrak{a}}{\mathfrak{b}}\,\frac{1}{1+\mathfrak{a}^2/\mathfrak{b}^2}\,.
\end{equation}
In order to understand this result, we need to consider the background Einstein's equation:
\begin{equation}
    f\, \left(u^2\, {c'}^2+12\right)+4 \left(m^2\,W-u \,f'-3 \right)\,=\,0\,,
\end{equation}
and solve it close to the black hole horizon $u=u_h$. Then we have:
\begin{equation}
   -4\pi\, u_h\,T\,e^{\chi(u_h)/2}\,+\,3\,-\,m^2\,u_h^{2\mathfrak{b}}\,(\cosh{(\Omega\,-\,c_h)})^{\mathfrak{a}}\,=\,0\,.\label{c1}
\end{equation}
One can show that the first term tends to a constant value at large strain (see Fig.~\ref{checks}) and it can be therefore discarded at least for the scaling analysis. Therefore, we obtain that:
\begin{equation}
    s\,\sim\,(\cosh{(\Omega-c_h)})^{\mathfrak{a}/\mathfrak{b}}\,,
\end{equation}
which in the limit $\Omega \gg c_h$ reduces to the simple scaling:
\begin{equation}
    s\,\sim\,\varepsilon^{2\,\mathfrak{a}/\mathfrak{b}}\,.\label{s2}
\end{equation}
We numerically verify (see Fig.~\ref{checks}) that $\Omega \gg c_h$ only when $\mathfrak{a}\ll \mathfrak{b}$, which is exactly the regime in which our analytic formula Eq.~\eqref{s2} works. More generally, a correction has to be taken into account and the final scaling is given by Eq.~\eqref{escal} as shown in Fig.~\ref{nonl}(B).

Combing the two results { Eq.~\eqref{scalfin}} and Eq.~\eqref{escal}, we can predict a universal relation between the stress and the entropy given by:
\beq
s \sim \sigma^{\xi},\quad \text{with}\quad  \xi^{-1}\,=\,\frac{3}{2}\,\left(1+\frac{\nu^2}{9}\right)\,, \label{fi}
\eeq
where $\nu$ is the non-linear shear scaling in Eq.~\eqref{scalfin}. This prediction is numerically confirmed as displayed in Fig.~2(C). Notice that similar scalings between the non-linear stress and the external strain can be derived withing the EFT formalism of~\cite{PhysRevD.102.069901}. Nevertheless, such a field theory construction does not provide a direct computation for the entropy of the system.

\section{Low temperature entropy from the gravitational model}
\begin{figure*}[h]
\begin{center}
\includegraphics[width=0.32\linewidth]{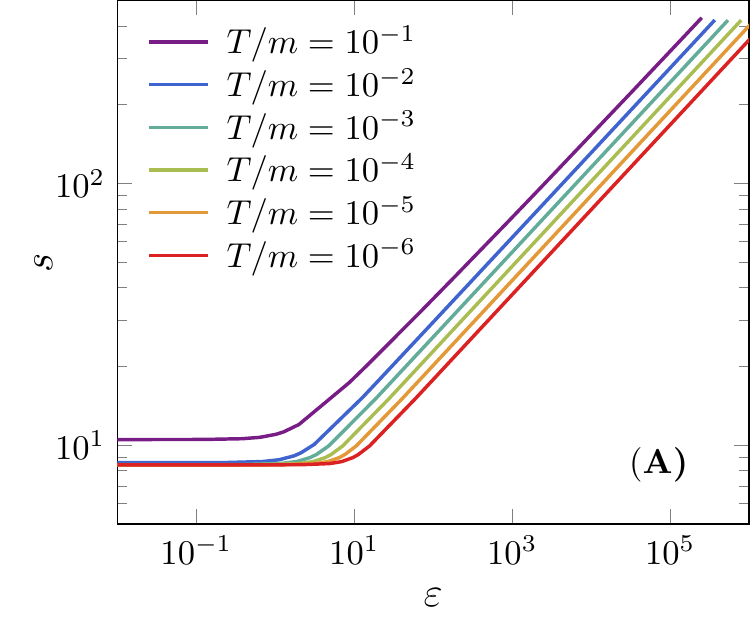}
\includegraphics[width=0.32\linewidth]{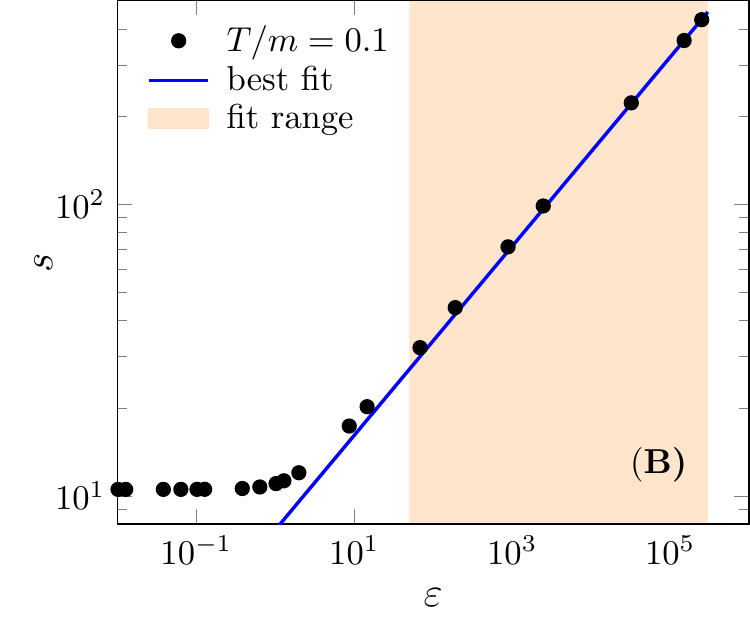}
\includegraphics[width=0.33\linewidth]{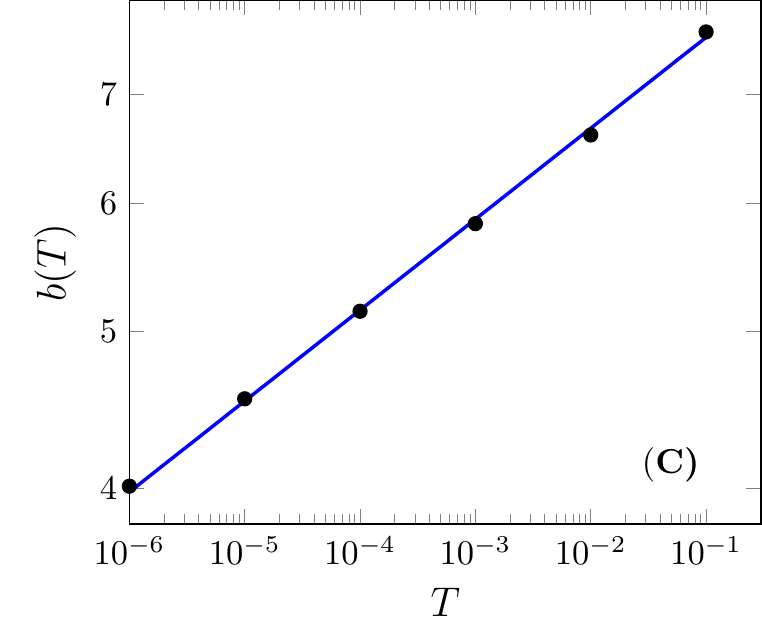}
    \caption{\textbf{Low temperature entropy at finite shear strains in the gravity theory.} \textbf{(A)} Entropy density-strain curves for different temperatures. \textbf{(B)} Fit of the entropy density-strain data in the large strain region to $s=b(T) \epsilon^{\zeta}$ (where $T/m=0.1$). \textbf{(C)}  The function $b(T)$ with respect to the temperature $T$. The best fit gives $b(T)\sim4.3 T^{0.06}$. The potential is fixed to $ W(\mathcal{X},\mathcal{Z})\,=\,\mathcal{X}^{\frac{11}{24}}\,\mathcal{Z}^{\frac{55}{48}}$.}.\label{fig:LTE}
\end{center}
\end{figure*}

In order to check the behavior of  entropy in the gravitational description
as the temperature $T\rightarrow0$, we plot the entropy with respect to shear strain at low temperatures, as shown in Fig.~\ref{fig:LTE}. We find that the entropy at large shear strains is parametrized as $s=b(T) \epsilon^{\zeta}$ with $\zeta$ the power defined in Eq.~\eqref{escal} and the residual zero temperature constant term is negligible in that limit.

By fitting the numerical data at large strain and small temperature, we obtain that $b(T)\sim4.3 T^{0.06}$. This outcome is consistent with the arguments in \cite{Baggioli2020} about an emergent Lifshitz-like anisotropic geometry in the limit of large strain.

In the case of amorphous solids, the zero-temperature limit corresponds to the regime where the temperature is small compared with any other energy scales, but obviously not exactly zero. Mapping this situation to our holographic setup, we should therefore consider the black hole at sufficiently low but finite temperature. Indeed, as we show in Fig.~\ref{fig:LTE}, the entropy for $T\ll 1$ remains non-zero, resembling a key feature of amorphous solids. Moreover, the scaling (for large $\sigma$),
\beq
s \sim \sigma^{\xi}\,,
\eeq
is universal at sufficiently low temperatures. Therefore, the entropy increases under shear, according to our gravitational theory.

\section{Additional simulation data for the frictionless granular model in 3D}
\label{addi}

\subsection{Discussion on the power-law fitting of stress-strain curves in the shear hardening regime}
\begin{figure}[ht!]
    \centering
    \includegraphics[width=\linewidth]{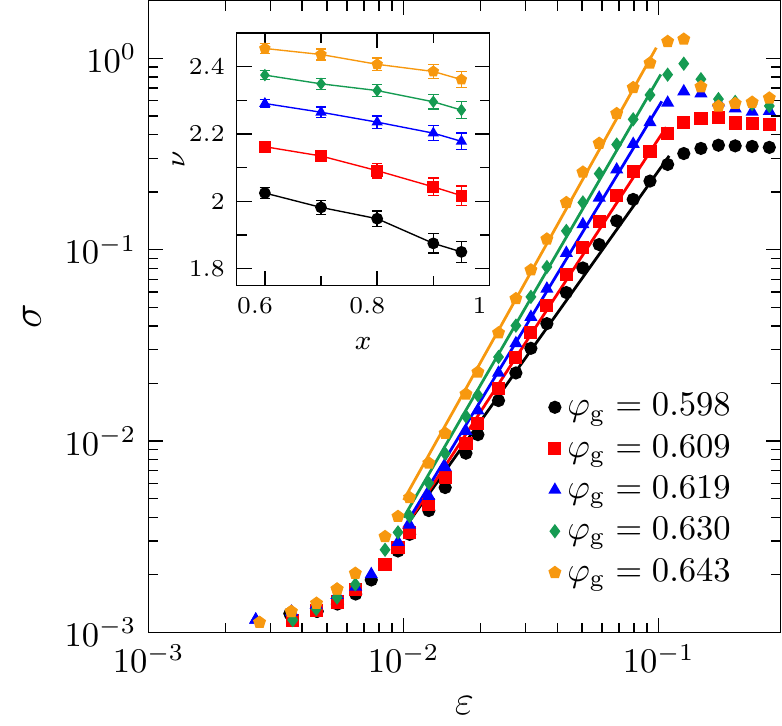}
     \caption{
     \textbf{Power-law fitting of the shear hardening part on stress-strain curves.}
    Data are obtained from simulations of the 3D frictionless model,  for $P_{0} = 10^{-2}$ and a few different $\varphi_{\rm g}$, on a log-log scale. The inset shows the dependence of the non-linear exponent $\nu$ on the parameter $x$ and the error bars represent the standard error of the fitting coefficient. }
    \label{fig:loglog_p1ei5}
\end{figure}

To show more clearly the power law behavior of the stress-strain curves in Fig.~3(A), we plot the data on a  log-log scale  (Fig.~\ref{fig:loglog_p1ei5}).
As can be seen from the plot, the data in the hardening regime, which begins around $\varepsilon = 10^{-2}$ and ends around yielding, can be nicely fitted by straight lines on the log-log scale. More precisely, we fit the data to the power-law scaling,   $\sigma \sim \epsilon^{\nu}$, in the window $\epsilon \ge 10^{-2}$ and $\sigma < x\, \sigma_{\rm max}$, where $\sigma_{\rm max}$ is the maximum stress and $x$ is a predetermined parameter.  The $x$-dependence of the exponent $\nu$ is shown in the inset of Fig.~\ref{fig:loglog_p1ei5}.
The exact value of $\nu$ depends on the choice of $x$, but the trend is robust (in Fig.~3(A), we set $x = 0.9$).

\subsection{Estimation of the yielding stress and strain}

\begin{figure}[ht!]
    \centering
    \includegraphics[width=\linewidth]{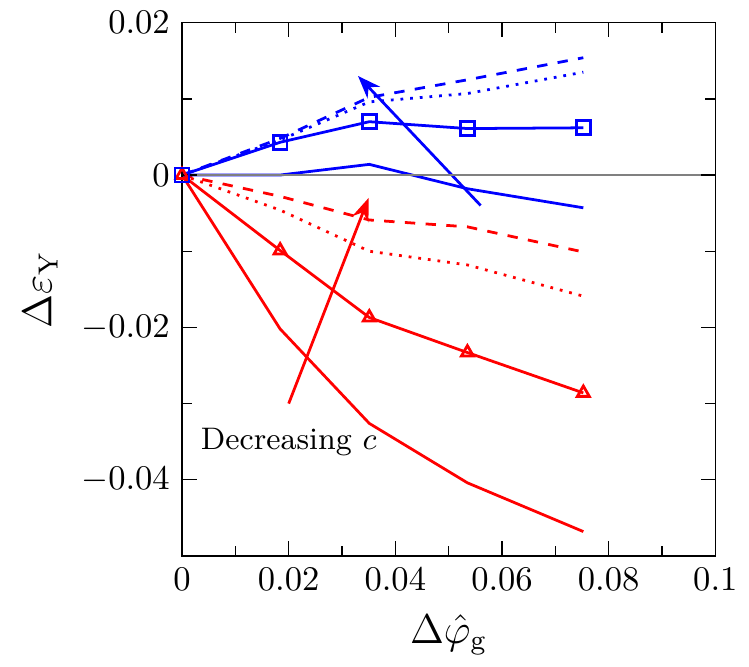}
     \caption{
     \textbf{Dependence of the yielding strain on the parameter $c$.}
    The yielding strain difference $\Delta \epsilon_{\rm Y}$ is plotted as a function of re-scaled degree of annealing $\Delta \hat{\varphi}_{\rm g}$,
     for $c=1, 0.98, 0.90, 0.80$ (3D frictionless soft sphere (SS) model).
     \textbf{Red}   color for $P_0 = 10^{-2}$ and \textbf{blue} for $P_0 = 10$.
      }
    \label{fig:yieldDef}
\end{figure}

The yielding point $\{ \epsilon_{\rm Y}, \sigma_{\rm Y}\}$ is estimated  at $\sigma_{\rm Y} = c \sigma_{\rm max}$.
The maximum stress $\sigma_{\rm max}$ is determined from a high order polynomial fit of the stress-strain peak. To test the robustness of our results on the parameter $c$, we plot the estimated $\Delta \epsilon_{\rm Y} = \epsilon_{\rm Y}(\Delta \hat{\varphi}_{\rm g}) - \epsilon_{\rm Y}(0)$ as a function of the  re-scaled degree of annealing $\Delta \hat{\varphi}_{\rm g}$, for several different $c$ (Fig.~\ref{fig:yieldDef}). 
Similar to Fig.~3 (where $c = 0.98$), $\epsilon_{\rm Y}$ decreases
with $\Delta \hat{\varphi}_{\rm g}$ for $P_0 = 10^{-2}$ and
increases (or
remains nearly constant) for $P_0 = 10$.
Note that $c=1$ is not ideal due to the ambiguity in determining the peak of the stress-strain curve, for the case of $\varphi_{\rm g} = 0.598$ (see Fig.~3(A)), where the curve is nearly flat after yielding.

{\subsection{2D nature of simple shear}}

\begin{figure}[ht!]
    \centering
    \includegraphics[width=\linewidth]{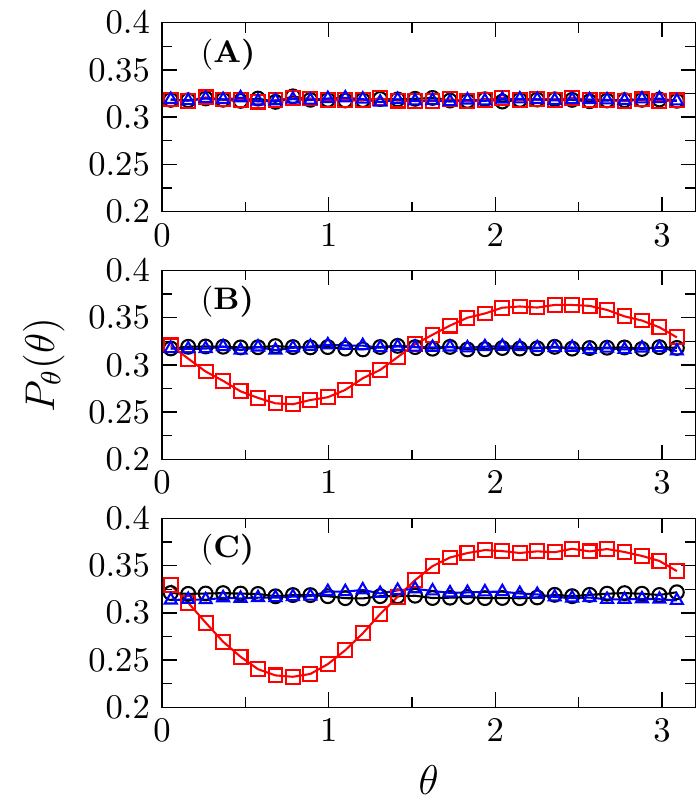}
     \caption{{\textbf{Distributions of projected contact angles under simple shear.}
     We plot the distributions   of projected contact angles
     in $x-y$ (square), $x-z$ (triangle) and $y-z$ (circle) planes, at three different shear strains (A) $\varepsilon = 0$, (B) $\varepsilon = 0.05$  and (C) $\varepsilon = 0.1$. The simple shear is applied in the  $x-y$ plane for 3D frictionless systems with $\varphi_{\rm g} = 0.643$.}  }
    \label{fig:2dNature}
\end{figure}

In this subsection, we verify that the effective rheology of our 3D system under simple shear deformations is  two-dimensional. Therefore, it is sufficient to consider a 2D theory as presented above.
In our 3D simulations, the system is deformed by simple shear in the $x-y$ plane. We compute the distributions of projected contacting angles between particles in all the three planes, at different shear strains (see Fig.~\ref{fig:2dNature}). The unstrained systems are isotropic as expected: the contacting angles distribute evenly in all three planes. As the system is strained, the distribution $P_\theta(\theta_{xy})$ in the shear plane becomes anisotropic, while in other two planes the distributions $P_\theta(\theta_{xz})$ and $P_\theta(\theta_{yz})$ remain isotropic. This proves that the dynamics of the system under shear is non-trivial only in the $x-y$ plane and that the third dimension can be safely neglected in the theory.

{ \subsection{Breaking point in the single stress-strain curve}}
\begin{figure}[ht!]
    \centering
    \includegraphics[width=\linewidth]{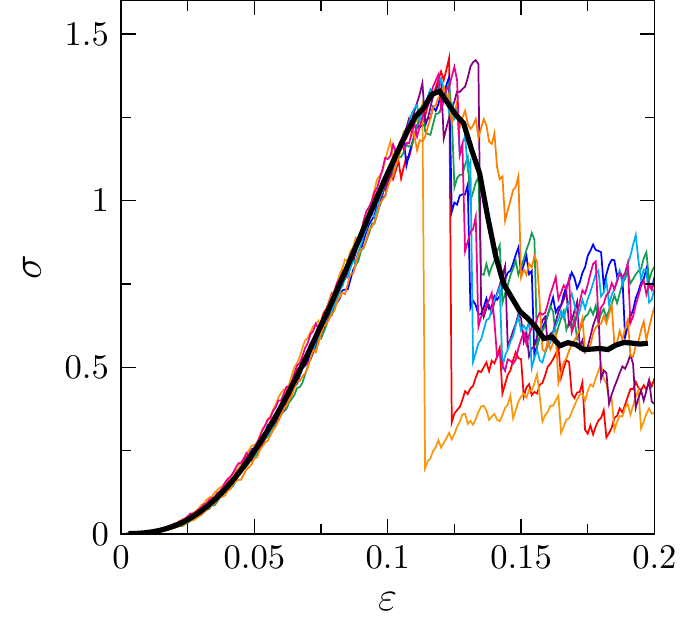}
     \caption{{
     \textbf{Single and average stress-strain curves.}
     Single (thin lines) and average (black bold line) stress-strain curves for the 3D frictionless  model with  $\varphi_{\rm g} = 0.643$.}}
    \label{fig:singleAve}
\end{figure}

The response of amorphous solid to quasi-static external strain is not purely elastic since it contains a non-negligible amount of plasticity\cite{biroli2016breakdown}. The {\it single strain-stress curve} of an individual simulated sample is not continuous but presents several local stress drops due to plasticity, as shown in Fig.~\ref{fig:singleAve}.
Furthermore, for deeply annealed samples, yielding can be defined as the breaking point at which the stress presents a sharp drop signaling the onset of a global instability.
This instability is related to the breakdown of the solid elastic behaviour due to plasticity -- after yielding the system does not react anymore as a solid. The definition of breaking point used in our theoretical framework is equivalent to
this criterion.

However, the yielding point defined in this way varies among samples. To suppress the sample-to-sample fluctuations, we perform averaging over the independent samples. When the {\it average stress-strain curve} is considered, yielding can be practically defined as the point at which the derivative of the stress vanishes. As evident from Fig.~\ref{fig:singleAve}, the location of the global instability in the single samples and the location of the yielding point defined as $d\sigma/d\epsilon=0$ in the average curve are close. This is also consistent with the theoretical argument presented in Sec.~S2, where sample-to-sample fluctuations are neglected. Note that all simulation data presented in this study refer to the averaged values unless otherwise specified.

\subsection{Discussion on the configurational entropy}
\begin{figure}[ht!]
    \centering
    \includegraphics[width=\linewidth]{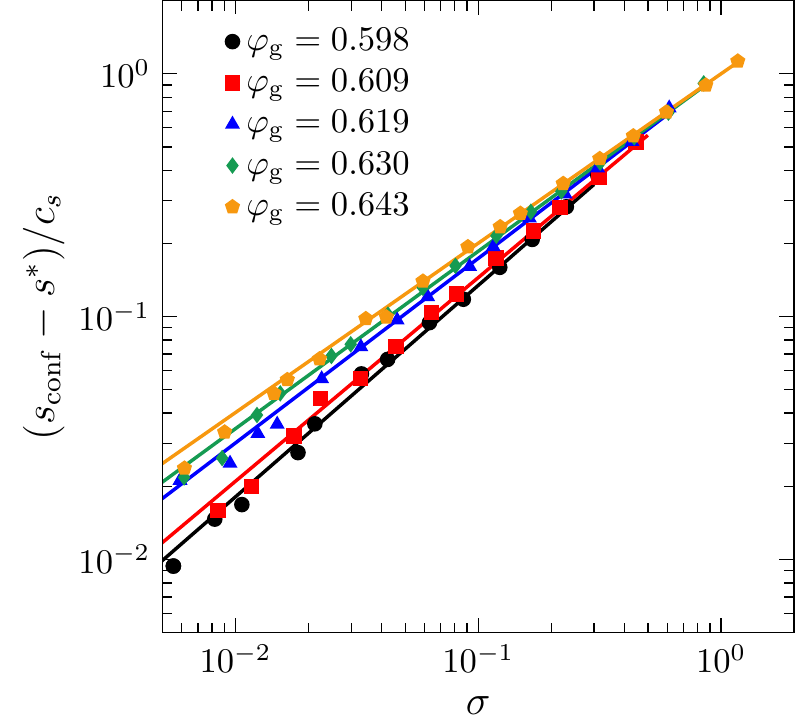}
     \caption{
     \textbf{Power-law fitting of the configurational entropy obtained in simulations.}
     Rescaled configurational entropy as a function of $\sigma$, on a log-log scale (3D frictionless model, $P_0=10^{-2}$).
     The dashed lines present the  fitting, $s_{\rm conf} = s^{*} + c_s \sigma^{\xi}$, where $s^{*}$, $c_s$ and $\xi$ are fitting parameters.
      }
    \label{fig:loglog_sconf}
\end{figure}

The configurational entropy is obtained in the following way.
A strained configuration at $\epsilon$ (for $\epsilon < \epsilon_{\rm Y}$) is
quasi-statically decompressed (keeping the strain unchanged) until it unjams at $\varphi_{\rm j}(\epsilon)$.
The decompression step size is $\delta \varphi = -10^{-4}$ for $P \ge 1.0$, $\delta \varphi = -2.5\times10^{-5}$ for $0.3 \le P <1.0$, $\delta \varphi = -5\times10^{-6}$ for $0.03 \le P < 0.3$, and $\delta \varphi = -10^{-6}$ for $P < 0.03$. The configuration entropy data  $s_{\rm conf}(\varphi_{\rm g})$ are collected from Ref.~\cite{berthier2017configurational}, and the relationship $\varphi_{\rm j}(\varphi_{\rm g})$ from Ref.~\cite{jin2021jamming}, for the same model. Together with the stress-strain curve $\sigma(\epsilon)$, we obtain $s_{\rm conf}(\sigma)$ (see Fig.~4(B)).

As suggested by Eq.~(3), the entropy data in Fig.~4(B) are fitted to $s_{\rm conf} = s^{*} + c_{s} \sigma^{\xi}$.
The rescaled plot of $(s_{\rm conf} - s^{*})/c_{s}$ versus $\sigma$ is presented in  Fig.~\ref{fig:loglog_sconf}, on a log-log scale (same data as in Fig.~4(B)).

\subsection{Edwards entropy}

\begin{figure*}[!ht]
    \centering
        \includegraphics[width=\linewidth]{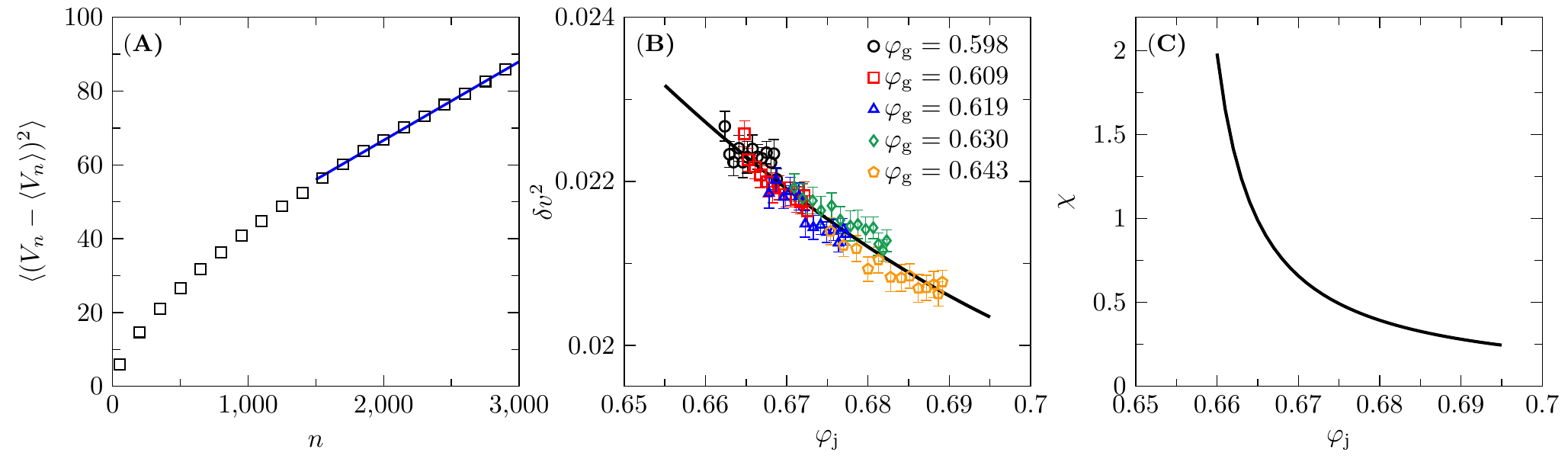}
    \caption{
    \textbf{Edwards entropy.}
    (A) Volume fluctuation $\langle (V_n - \langle V_n \rangle)^2 \rangle$ as a function of the cluster size $n$ in a typical example.
    The line represents a linear fit for $n \geq 1500$, in order to obtain $\delta v^2 \equiv \frac{\langle (V_n - \langle V_n \rangle)^2 \rangle}{n}$.
    (B) Intensive volume fluctuation $\delta v^2$ versus the jamming density $\varphi_{\rm j}$, for a few different $\varphi_{\rm g}$. The solid line represents the quadratic fitting $\delta v^2(\varphi_{\rm j}) = 0.3173 - 0.8057 \varphi_{\rm j} + 0.5445 \varphi_{\rm j}^2 $.
    The error bars represent the standard error for the fitted parameters.
    (C)  Compactivity $\chi$ versus the jamming density $\varphi_{\rm j}$, computed based on the quadratic fitting in (B).}
    \label{fig:ed_entropy}
\end{figure*}

Edwards statistical mechanics of granular matter was introduced by Edwards and co-workers\cite{edwards1989theory,RevModPhys.90.015006}.
Because energy is not conserved in granular matter, the choice of the standard microcanonical or canonical ensembles is not applicable.
Edwards et al.~\cite{edwards1989theory} proposed an alternative volume ensemble, based on which  the  {\it Edwards entropy} can be defined as the logarithm of the number of micro-states for a given volume.
The Edwards entropy in athermal granular matter plays a role analogous to the thermodynamic entropy in thermal systems.

Here, following a previously proposed method\cite{PhysRevLett.101.188001}, we calculate the Edwards entropy of our granular model under simple shear. This method has been used for isotropic systems without shear, in both simulation\cite{PhysRevLett.101.188001,PhysRevE.80.031301, jin2010first} and experimental studies\cite{PhysRevLett.127.018002}.
In order to apply the method to sheared configurations, the Lees-Edwards boundary conditions~\cite{lees1972computer} are used.
The calculation is  based on an analysis of the volume fluctuations through a generalized fluctuation-dissipation relation. The computation is performed for systems at the jamming density $\varphi_{\rm j}(\varepsilon)$ (where $P \approx 0$, see Fig. 4(A)), which are decompressed from sheared configurations $\varphi(\varepsilon)$ ($P > 0$) with the strain $\epsilon$ fixed.
We describe below the procedure and specify quantities that need to be calculated.

\textbf{(i) Volume fluctuations.} The local Voronoi volume of each particle is defined through the radical Voronoi tessellation, realized  using the Voro++ library\cite{voro++}.
One then calculates the total volume $V_n$ of a cluster formed by $n$ particles surrounding a reference particle, and the variance $\langle (V_n - \langle V_n \rangle)^2 \rangle$, where $\langle \cdots \rangle$ represents the average over configurations.
We find that for $n \geq 1500$, the variance $\langle (V_n - \langle V_n \rangle)^2 \rangle$ scales linearly with $n$ as shown in Fig.~\ref{fig:ed_entropy}(A), which means that the volume fluctuation is extensive in the large-$n$ limit and an intensive quantity $\delta v^2 \equiv \frac{\langle (V_n - \langle V_n \rangle)^2 \rangle}{n}$ can be defined. Interestingly, Fig.~\ref{fig:ed_entropy}(B) shows that the data of $\delta v^2$ versus $\varphi_{\rm j}$ collapse onto a master curve for different $\varphi_{\rm g}$ representing different degrees of annealing. This evidence justifies the application of Edwards theory: even though jammed states are protocol-dependent ($\varphi_{\rm g}$-dependent), they can be unified in a single volume ensemble for the evaluation of Edwards entropy. The data are then fitted with a quadratic function to obtain $\delta v^2(\varphi_{\rm j})$ (see Fig.~\ref{fig:ed_entropy}(B)).

\textbf{(ii) Compactivity.}
The {\it compactivity} $\chi$, which plays the role of ``temperature" in the volume ensemble,  is defined via the generalized fluctuation relation (we have set the ``Boltzmann constant" $\lambda$ to one),
\beq
\langle (V_n - \langle V_n \rangle)^2 \rangle =  \frac{\chi^2 d\langle V_n \rangle}{d\chi},
\eeq
or via the equivalent integral form,
\beq
\frac{1}{ \chi (\varphi_{\rm j})}  = \int_{\varphi_{\rm r}}^{\varphi_{\rm j}} \frac{1}{\varphi^2 \delta v^2(\varphi)}  d\varphi\,.
\label{eq:fluctuation2}
\eeq
Here, $\varphi = \frac{n V_{\rm g}}{\langle V_n \rangle}$ and we have set the average volume of grains $V_{\rm g} = 1$ in Eq.~(\ref{eq:fluctuation2}). In addition, we have chosen the lowest jamming  density  (the J-point density $\varphi_{\rm J}$~\cite{o2003jamming}) as reference, $\varphi_{\rm r} = \varphi_{\rm J} \approx 0.655$~\cite{jin2021jamming}, whose compactivity has been set to infinity, $\chi_{\rm r} = \infty$. Based on Eq.~(\ref{eq:fluctuation2}) and the quadratic fitting of $\delta v^2(\varphi_{\rm j})$ in Fig.~\ref{fig:ed_entropy}(B), we obtain $\chi(\varphi_{\rm j})$ (see Fig.~\ref{fig:ed_entropy}(C)).\\

\textbf{(iii) Edwards entropy.} Finally, the Edwards entropy $s_{\rm Ed}(\varphi_{\rm j})$ (per grain) is given by
\beq
\chi^{-1} = -\varphi_{\rm j}^2 \frac{d s_{\rm Ed}} {d \varphi_{\rm j}},
\eeq
or equivalently,
\beq
s_{\rm Ed}(\varphi_{\rm j}) &=& s_{\rm Ed}(\varphi_{\rm r}) - \int_{\varphi_{\rm r}}^{\varphi_{\rm j}} \frac{1}{\varphi^2 \chi(\varphi)} d\varphi\\
&=& s_{\rm Ed}(\varphi_r) + \Delta s_{\rm Ed}(\varphi_{\rm j}).
\eeq
The relevant quantity is the entropy change, $\Delta s_{\rm Ed}(\varphi_{\rm j}) = - \int_{\varphi_{\rm r}}^{\varphi_{\rm j}} \frac{1}{\varphi^2} \frac{1}{\chi(\varphi)} d\varphi$.
Here, $\Delta s_{\rm Ed}(\varphi_{\rm j})$ is negative, which is reasonable since we expect that $s_{\rm Ed}(\varphi_{\rm j}) < s_{\rm Ed}(\varphi_{\rm r})$ for $\varphi_{\rm j} > \varphi_{\rm r}$ (the number of possible packings should decrease with the packing density). Because $\varphi_{\rm j}(\epsilon)$ and $\sigma (\epsilon)$ are available (Fig.~4(A) and Fig.~3(A)),  $\Delta s_{\rm Ed}(\varphi_{\rm j})$ can be easily converted to a function of the stress, $\Delta s_{\rm Ed}(\sigma)$ (see Fig.~4(C)).

As shown in Fig.~4(C), the Edwards entropy $\Delta s_{\rm Ed}(\sigma)$ increases with the stress $\sigma$, consistent with the data of configurational entropy in Fig.~4(B). As suggested by Eq.~(3), we fit the data to a power law, $\Delta s_{\rm Ed} = \Delta s^{*}_{\rm Ed} + c_{\rm Ed} \sigma^{\xi_{\rm Ed}}$, where $\Delta s^{*}_{\rm Ed}$, $c_{\rm Ed}$ and $\xi_{\rm Ed}$ are fitting parameters. The obtained exponent $\xi_{\rm Ed}$ as a function of stress exponent $\nu$ is plotted in the inset of Fig.~4(C), together with the exponent $\xi$ obtained from the configurational entropy in Fig.~4(B). Both $\xi$ and $\xi_{\rm Ed}$  decrease with $\nu$ as predicted by our theory.

\subsection{Reversibility of shear hardening}
\begin{figure}[ht!]
    \centering
    \includegraphics[width=\linewidth]{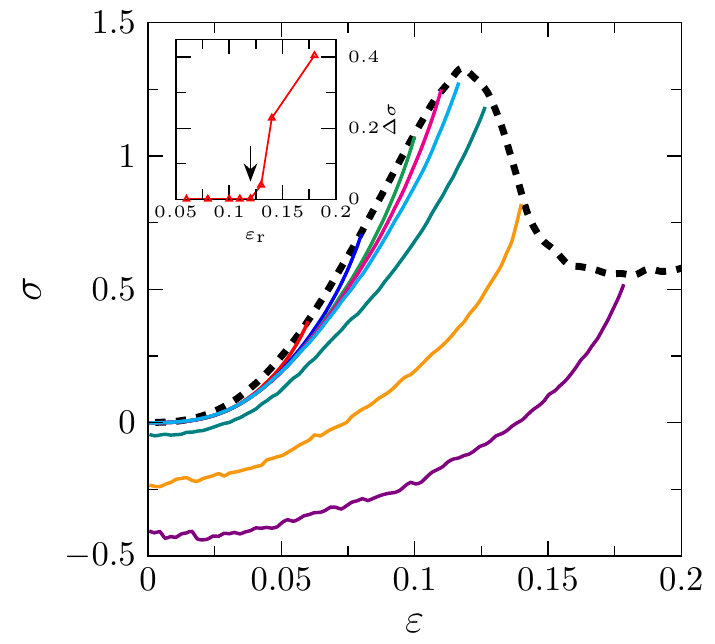}
     \caption{\textbf{Reversibility test of shear hardening.}
          Loading and unloading curves of the 3D frictionless granular model, for $P_{0}=10^{-2}$ and  $\varphi_{\rm g} = 0.643$. The shear strain is applied up to a reserve strain $\epsilon_{\rm r}$ (dashed line), and then is reversed (solid lines), where $\epsilon_{\rm r} = 0.06, 0.08, 0.10, 0.11, 0.12, 0.13, 0.14, 0.18$ (from left to right).
     The stress peak appears around $\epsilon_{\rm Y} = 0.12$.
     The inset shows the dependence of the residual stress for one cycle, $\Delta \sigma = \sigma_{\rm before}(\epsilon=0) -  \sigma_{\rm after}(\epsilon=0)$, on $\epsilon_{\rm r}$,  where $\epsilon_{\rm Y}$ is indicated by the arrow.
     }
    \label{fig:reverse}
\end{figure}

Here we show that the shear hardening behavior observed in our simulations are primarily  elastic. Indeed, as presented in Fig.~\ref{fig:reverse}, the  stress-strain curve is nearly reversible before yielding~\cite{saw2016nonaffine}, $\epsilon_{\rm r} < \epsilon_{\rm Y}$, where $\epsilon_{\rm r}$ is the reverse strain. For comparison purposes, we also plot curves for  $\epsilon_{\rm r} > \epsilon_{\rm Y}$, which are clearly irreversible.  The cyclic shear test thus clearly demonstrates that plastic effects  have a minor impact on shear hardening.

\subsection{Absence of shear hardening in over-compressed systems}
\begin{figure}[ht!]
    \centering
    \includegraphics[width=\linewidth]{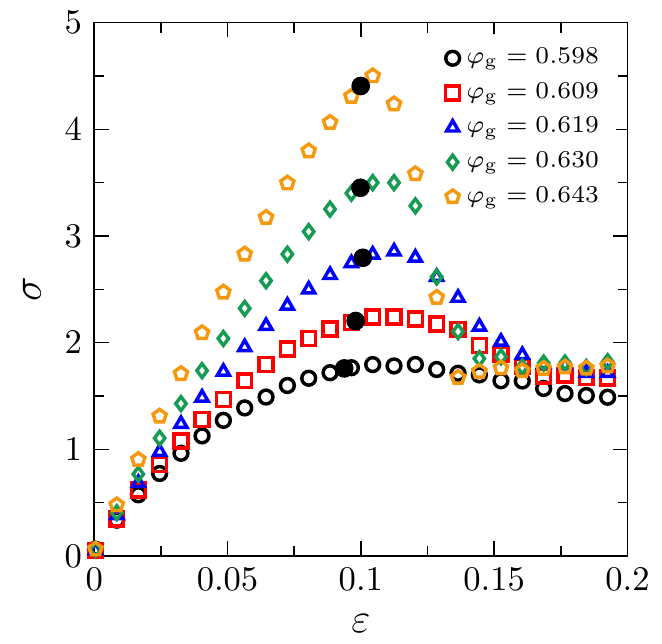}
     \caption{
     \textbf{Stress-strain curves of over-compressed systems.}
    Stress-strain curves obtained from initially over-compressed 3D frictionless systems ($P_{0}=10$), for  a few different $\varphi_{\rm g}$. The yielding point $\{ \epsilon_{\rm Y}, \sigma_{\rm Y}\}$ is estimated  at $\sigma_{\rm Y} = c \sigma_{\rm max}$, where $\sigma_{\rm max}$ is the maximum stress and $c = 0.98$.}
    \label{fig:stress_strain_p1ei2}
\end{figure}

Stress-strain curves for initially over-compressed systems ($P_{0}=10$) are presented in Fig.~\ref{fig:stress_strain_p1ei2}.
Contrary to the case of  $P_0 = 10^{-2}$ (see Fig.~3(A)), shear hardening disappears, and correspondingly the yielding strain $\epsilon_{\rm Y}$ does not decrease with $\varphi_{\rm g}$ anymore (see Fig.~3(C)).
When $P_0 \to 0$, the coordination number $Z$ (average number of contacts on each particle) approaches $2d$ with $d$ the spatial dimension. The above results suggest that, the isostatic condition, $Z = 2d$, is essential for shear hardening in our systems.

\subsection{Shear hardening in mechanically trained systems}
The simulation results presented in the main text are obtained from thermally annealed systems, by using the swap algorithm. Here we present additional data   from mechanically annealed systems ($N = 2000$), using athermal quasi-static cyclic shear~\cite{babu2021dilatancy}.
The procedure is as follows: (i) Firstly, random configurations are generated and slowly compressed to a packing fraction $\varphi_0 = 0.66$, slightly above the J-point density (minimum jamming density) $\varphi_{\rm J} = 0.655$~\cite{jin2021jamming}.
(ii) Next, we apply cyclic shear under constant volume, athermal, quasi-static conditions, with a step size $\delta \epsilon = 10^{-3}$. During one cycle, the shear strain $\epsilon$ is changed as $\{0 \to \epsilon_{\rm max} \to -\epsilon_{\rm max} \to 0 \}$, where $\epsilon_{\rm max} = 0.05$.
(iii)
The cycles  stop if the system remains unjammed (the energy per particles is less than $10^{-13}$) during one entire cycle. This step increases the jamming density of the system from $\varphi_{\rm j} = \varphi_{\rm J} = 0.655$ to $\varphi_{\rm j} = 0.662$.
(iv) The unjammed configuration at $\{\varphi_0 = 0.66, P_0=0\}$ is slowly compressed to a target pressure $P_0 = 10^{-2}$.

\begin{figure}[ht!]
    \centering
    \includegraphics[width=\linewidth]{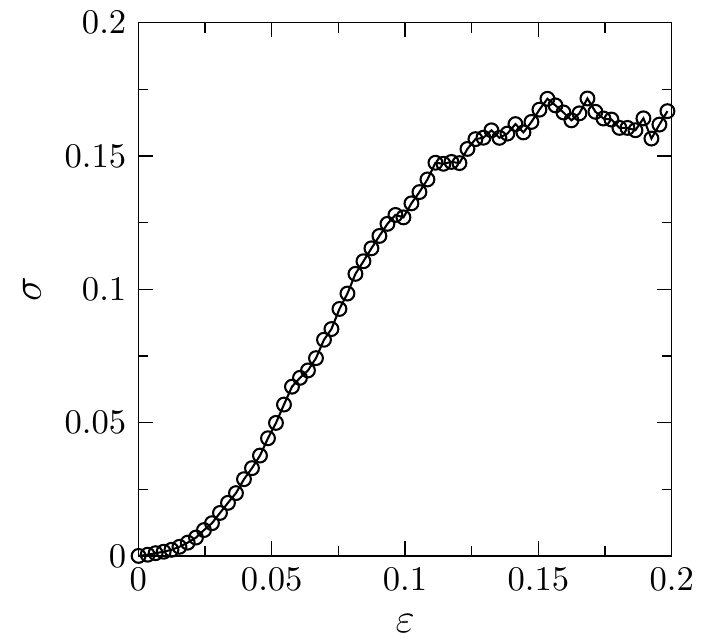}
     \caption{\textbf{
     Stress-strain curve of mechanically trained systems.}
    Stress-strain curve of a 3D frictionless system annealed by cyclic shear ($P_0 = 10^{-2}$). }
    \label{fig:annealCycShear_p1ei5}
\end{figure}

The above procedure prepares unstrained initial configurations. A stress-strain curve for such a configuration is plotted in Fig.~\ref{fig:annealCycShear_p1ei5}, which shows clearly a shear hardening effect.
It is known that mechanical annealing is less efficient compared to swap annealing~\cite{babu2021dilatancy}. As a result, shear hardening  is also weaker in mechanical annealed systems. Nevertheless, we conclude that this effect is independent of the annealing protocol, as demonstrated here.

\section{Simulations of a frictionless granular model in 2D}

\begin{figure*}[ht!]
    \centering
    \includegraphics[width=\linewidth]{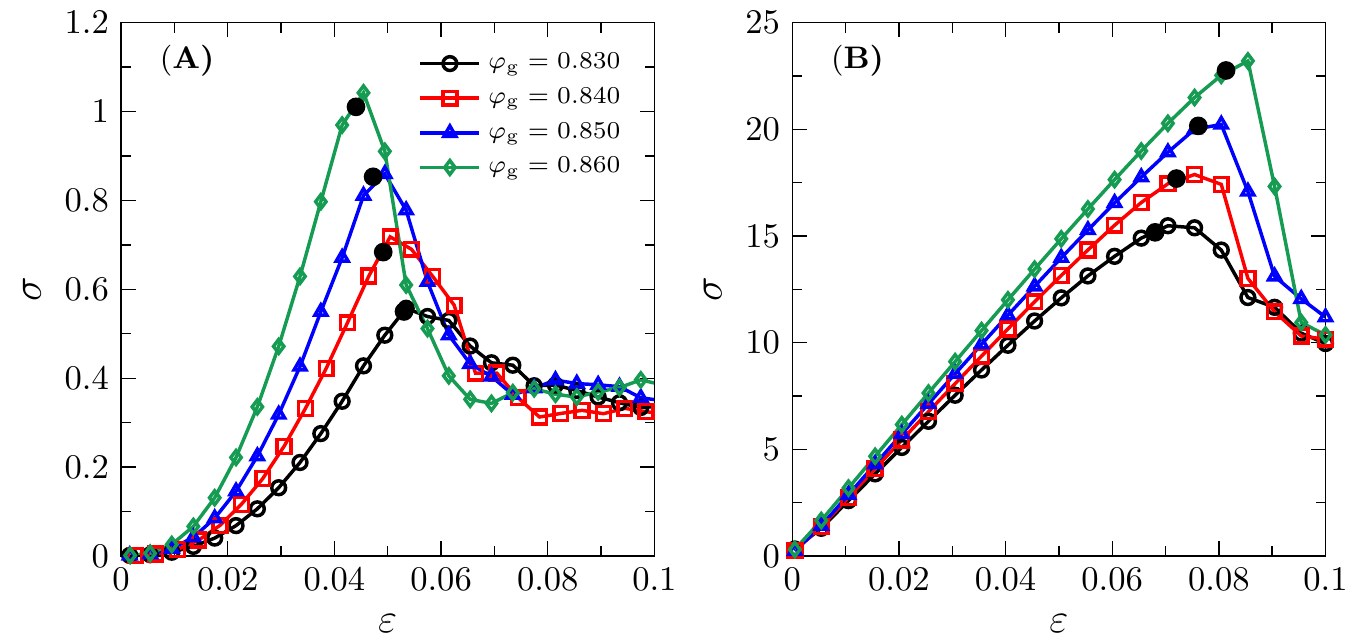}
     \caption{
     {\bf Stress-strain curves of 2D systems.}
     Stress-strain curves for \textbf{(A)} $P_0 = 10^{-2}$  and \textbf{(B)} $P_0 = 10^2$  of a 2D frictionless granular model. Yielding points $\{ \epsilon_{\rm Y}, \sigma_{\rm Y}\}$ are estimated  at $\sigma_{\rm Y} = c \sigma_{\rm max}$, where $\sigma_{\rm max}$ is the maximum stress and $c = 0.98$.}
    \label{fig:2dSim}
\end{figure*}

The 2D frictionless model is composed of $N = 4000$ soft disks, whose diameters are distributed according to  $P(D) \sim D^{-2}$, with $D_{\rm min} \le D \le D_{\rm min}/0.45$.
Two disks are in contact if their separation  $r_{kl}$ is less than their mean diameter $D_{kl} = (D_k + D_l)/2$.
Two contacting disks interact via a short-range repulsive  potential,
\beq
V(r_{kl}) = \frac{k_v}{2}\left(1-\frac{r_{kl}}{D_{kl}} \right)^2.
\eeq
The unit of length is the average diameter of all disks, the unit of energy is $10^{3}\times k_v$, and all disks have the same unit mass. The stress-strain curves (averaged over 24 independent samples) for $P_0 = 10^{-2}$ and $P_{0}=10^2$ are plotted in Fig.~\ref{fig:2dSim}, showing similar behavior as the case in 3D.

\section{Simulations of a frictional granular model in 2D}
\begin{figure}[ht!]
    \centering
    \includegraphics[width=\linewidth]{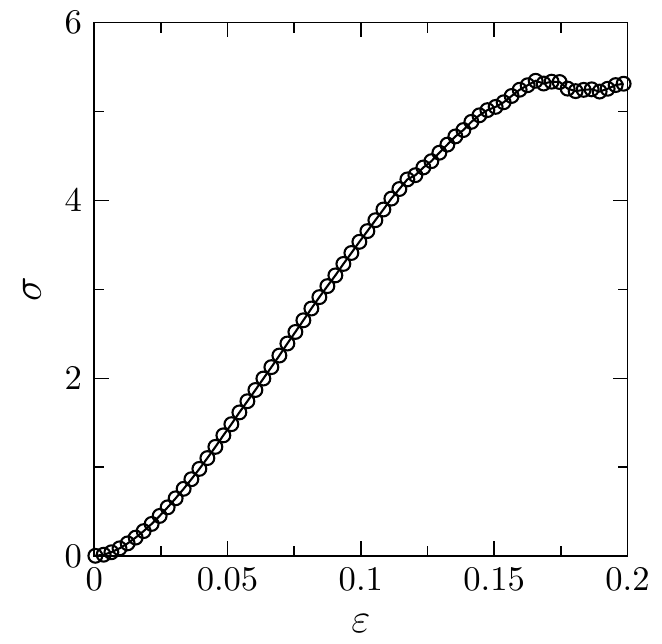}
     \caption{
     \textbf{Stress-strain curve of frictional systems.}
    We plot a stress-strain curve obtained from simulations of the 2D frictional granular model ($P_0 = 10^{-2}$).}
    \label{fig:2dFrictional_p1ei5}
\end{figure}

We simulate the same frictional 2D model as in Ref.~\cite{Otsuki2020Shear}. The force between two contacting particles is
\beq
\mathbf{F}_{kl} = \mathbf{F}_{kl}^n + \mathbf{F}_{kl}^t,
\eeq
where the normal component is
\beq
\mathbf{F}_{kl}^n = -k^n \left(r_{kl} - \frac{D_k + D_l}{2}\right) \mathbf{n}_{kl} - \zeta^n \mathbf{v}_{kl}^n.
\eeq
Here $\mathbf{n}_{kl}$ is the unit vector along the contact direction, $\mathbf{v}_{kl}^n = (\mathbf{v}_{kl} \cdot \mathbf{n}_{kl}) \mathbf{n}_{kl}$ the normal relative velocity, and $\mathbf{v}_{kl} = \mathbf{v}_k - \mathbf{v}_l$ the relative velocity. The magnitude of the tangential force has to satisfy the Coulomb condition,
\beq
| \mathbf{F}_{kl}^t | = \min\{ |\tilde{\mathbf{F}}_{kl}^t |, |\mu_{\rm f}  \mathbf{F}_{kl}^n|\},
\eeq
where $\mu_{\rm f}$ is the friction coefficient, and
\beq
\tilde{\mathbf{F}}_{kl}^t = -k^t \mathbf{u}_{kl}^t - \zeta^t \mathbf{v}_{kl}^t,
\eeq
with $\mathbf{u}_{kl}^t$ and $\mathbf{v}_{kl}^t$ being the tangential displacement and velocity. Both translational and rotational motions contribute to the tangential velocity at contact. Thus the tangential velocity is given by
\beq
\mathbf{v}_{kl}^t = \mathbf{v}_{kl} - \mathbf{v}_{kl}^n - \frac{D_k \omega_k + D_l \omega_l}{2},
\eeq
where $\omega$ is the angular velocity. The tangential displacement is  updated according to
\beq
\mathbf{\dot{u}}_{kl}^t = \mathbf{v}_{kl}^t,
\eeq
if $\|\tilde{\mathbf{F}}_{kl}^t\| \leq \|\mu \mathbf{F}_{ij}^n\|$, otherwise remains unchanged.

Our system contains  $N=4000$ grains, with equal numbers of larger and smaller grains. All grains have the same mass density, and the diameter ratio between big and small particles is 1.4. We set $k^t = 0.2 k^n, \zeta^t = \zeta^n = \sqrt{m_{\rm L} k^n}$ and $\mu_{\rm f} = 1.0$, where  $m_{\rm L}$ is the mass of larger grains. The units of length, mass and energy are $D_{\rm L}$, $m_{\rm L}$ and $10^3 \times k^n$.

Both compression (decompression) and shear are performed under athermal quasi-static conditions as in the frictionless case. The increment steps are $\delta \varphi = 10^{-4}$ and $\delta \epsilon = 10^{-4}$. Energy is minimized by following Newton's law, if $\sum_{k=1}^N \mathbf{v}_k \cdot \mathbf{F}_k \ge 0$, where $\mathbf{F}_k$ is the contacting force of particle $k$ (otherwise, both translational and rotational velocities are abandoned).
Mechanical equilibrium is reached when
the average force and the average torque per particle are both below $5\times 10^{-9}$.

We employ the cyclic shear method to obtain  annealed systems. Random configurations are generated at an area fraction $\varphi = 0.75$, and then slowly compressed to $\varphi = 0.845$.
The pressure becomes non-zero around $\varphi_{\rm J} = 0.829$, and thus the system is over-compressed at $\varphi = 0.845$. After 1000 cycles of shear with $\epsilon_{\rm max} = 0.04$, the jamming density increases to a larger value about $\varphi_{\rm j} = 0.838$. The configurations are then slowly decompressed to $P_0 = 10^{-2}$. The above procedure prepares annealed unstrained initial configurations. The shear hardening effect is not as strong as in Fig.~3(A),  but visible (see Fig.~\ref{fig:2dFrictional_p1ei5}). As shown in Fig.~3(A), this effect could be further magnified by increasing the degree of annealing.

\section{Proposal for experimental validations}
Finally, we expect  our numerical observations to be reproducible in experiments: (i) It is not always easy to avoid friction  in experimental systems, but our simulation result (Fig.~\ref{fig:2dFrictional_p1ei5}) suggests that friction does not destroy the non-linear shear hardening effect. (ii) Our simulation results suggest that shear hardening should be observable in both 2D disks~\cite{lechenault2008critical, bi2011jamming} and 3D spheres~\cite{kou2017granular}. (iii) It is possible to implement athermal quasi-static shear using the recently developed multi-ring Couette shear setup~\cite{zhao2019shear}. (iv) Mechanical annealing can be realized in experiments using cyclic shear{~\cite{zhao2022ultrastable,xing2021x}}. (v) Although direct estimation of the configurational entropy  is  difficult, one can measure the ``Edwards entropy" in experiments~\cite{PhysRevLett.127.018002}, based on a thermodynamic framework initially proposed by Edwards~\cite{edwards1989theory}.
%

 \balance

 \end{document}